\pdfoutput=1

\documentclass[12pt,a4paper]{article}

\usepackage{ifthen} 
\newboolean{pdflatex}
\setboolean{pdflatex}{true} 

\newboolean{articletitles}
\setboolean{articletitles}{true} 

\newboolean{uprightparticles}
\setboolean{uprightparticles}{false} 

\newboolean{inbibliography}
\setboolean{inbibliography}{false} 

\usepackage[top=1in, bottom=1.25in, left=1in, right=1in]{geometry}

\usepackage{microtype}
\usepackage{lineno}  
\usepackage{xspace} 
\usepackage{caption} 

\usepackage{graphicx}  
\usepackage{color}
\usepackage{colortbl}
\graphicspath{{./figs/}} 

\usepackage{amsmath} 
\usepackage{amssymb}
\usepackage{amsfonts}
\usepackage{upgreek} 

\newcommand*\patchAmsMathEnvironmentForLineno[1]{%
\expandafter\let\csname old#1\expandafter\endcsname\csname #1\endcsname
\expandafter\let\csname oldend#1\expandafter\endcsname\csname
end#1\endcsname
 \renewenvironment{#1}%
   {\linenomath\csname old#1\endcsname}%
   {\csname oldend#1\endcsname\endlinenomath}%
}
\newcommand*\patchBothAmsMathEnvironmentsForLineno[1]{%
  \patchAmsMathEnvironmentForLineno{#1}%
  \patchAmsMathEnvironmentForLineno{#1*}%
}
\AtBeginDocument{%
\patchBothAmsMathEnvironmentsForLineno{equation}%
\patchBothAmsMathEnvironmentsForLineno{align}%
\patchBothAmsMathEnvironmentsForLineno{flalign}%
\patchBothAmsMathEnvironmentsForLineno{alignat}%
\patchBothAmsMathEnvironmentsForLineno{gather}%
\patchBothAmsMathEnvironmentsForLineno{multline}%
\patchBothAmsMathEnvironmentsForLineno{eqnarray}%
}

\usepackage{hyperref}    
\usepackage[all]{hypcap} 

\usepackage{xspace} 
\usepackage{upgreek}

\def\lhcb {\mbox{LHCb}\xspace}

\def\MagUp {\mbox{\em Mag\kern -0.05em Up}\xspace}

\ifthenelse{\boolean{uprightparticles}}%
{

 \def\Ppsi        {\ensuremath{\uppsi}\xspace}

 \def\PDelta      {\ensuremath{\Delta}\xspace}                 
 \def\PXi      {\ensuremath{\Xi}\xspace}                 
 \def\PLambda      {\ensuremath{\Lambda}\xspace}                 
 \def\PSigma      {\ensuremath{\Sigma}\xspace}                 
 \def\POmega      {\ensuremath{\Omega}\xspace}                 
 \def\PUpsilon      {\ensuremath{\Upsilon}\xspace}                 
 
 \def\PB      {\ensuremath{\mathrm{B}}\xspace}                 
                  
 \def\PD      {\ensuremath{\mathrm{D}}\xspace}

 \def\PJ      {\ensuremath{\mathrm{J}}\xspace}                 
 \def\PK      {\ensuremath{\mathrm{K}}\xspace}

 \def\Pb      {\ensuremath{\mathrm{b}}\xspace}                 
 \def\Pc      {\ensuremath{\mathrm{c}}\xspace}

 \def\Pi      {\ensuremath{\mathrm{i}}\xspace}

 \def\Ps      {\ensuremath{\mathrm{s}}\xspace}

}
{

 \def\Ppsi        {\ensuremath{\psi}\xspace}                 
                  
 \mathchardef\PDelta="7101
 \mathchardef\PXi="7104
 \mathchardef\PLambda="7103
 \mathchardef\PSigma="7106
 \mathchardef\POmega="710A
 \mathchardef\PUpsilon="7107
                  
 \def\PB      {\ensuremath{B}\xspace}                 
                  
 \def\PD      {\ensuremath{D}\xspace}

 \def\PJ      {\ensuremath{J}\xspace}                 
 \def\PK      {\ensuremath{K}\xspace}

 \def\Pb      {\ensuremath{b}\xspace}                 
 \def\Pc      {\ensuremath{c}\xspace}

 \def\Pi      {\ensuremath{i}\xspace}

 \def\Ps      {\ensuremath{s}\xspace}

}

\makeatletter
\ifcase \@ptsize \relax
  \newcommand{\miniscule}{\@setfontsize\miniscule{4}{5}}
\or
  \newcommand{\miniscule}{\@setfontsize\miniscule{5}{6}}
\or
  \newcommand{\miniscule}{\@setfontsize\miniscule{5}{6}}
\fi
\makeatother

\DeclareRobustCommand{\optbar}[1]{\shortstack{{\miniscule (\rule[.5ex]{1.25em}{.18mm})}
  \\ [-.7ex] $#1$}}

\def\squark    {{\ensuremath{\Ps}}\xspace}

\def\cquark    {{\ensuremath{\Pc}}\xspace}

\def\bquark    {{\ensuremath{\Pb}}\xspace}

  \def\Kbar    {{\kern 0.2em\overline{\kern -0.2em \PK}{}}\xspace}

\def\KorKbar    {\kern 0.18em\optbar{\kern -0.18em K}{}\xspace}

  \def\Dbar    {{\kern 0.2em\overline{\kern -0.2em \PD}{}}\xspace}
\def\D       {{\ensuremath{\PD}}\xspace}

\def\DorDbar    {\kern 0.18em\optbar{\kern -0.18em D}{}\xspace}

\def\B       {{\ensuremath{\PB}}\xspace}
\def\Bbar    {{\ensuremath{\kern 0.18em\overline{\kern -0.18em \PB}{}}}\xspace}

\def\BorBbar    {\kern 0.18em\optbar{\kern -0.18em B}{}\xspace}
\def\Bz      {{\ensuremath{\B^0}}\xspace}
\def\Bzb     {{\ensuremath{\Bbar{}^0}}\xspace}
\def\Bu      {{\ensuremath{\B^+}}\xspace}
\def\Bub     {{\ensuremath{\B^-}}\xspace}
\def\Bp      {{\ensuremath{\Bu}}\xspace}
\def\Bm      {{\ensuremath{\Bub}}\xspace}

\def\Bsb     {{\ensuremath{\Bbar{}^0_\squark}}\xspace}

\def\jpsi     {{\ensuremath{{\PJ\mskip -3mu/\mskip -2mu\Ppsi\mskip 2mu}}}\xspace}

  \def\Y#1S{\ensuremath{\PUpsilon{(#1S)}}\xspace}

\def\Xires       {{\ensuremath{\PXi}}\xspace}

\def\Lz          {{\ensuremath{\PLambda}}\xspace}
\def\Lbar        {{\ensuremath{\kern 0.1em\overline{\kern -0.1em\PLambda}}}\xspace}
\def\LorLbar    {\kern 0.18em\optbar{\kern -0.18em \PLambda}{}\xspace}

\def\Omegares    {{\ensuremath{\POmega}}\xspace}

\def\Lb      {{\ensuremath{\Lz^0_\bquark}}\xspace}
\def\Lbbar   {{\ensuremath{\Lbar{}^0_\bquark}}\xspace}
\def\Lc      {{\ensuremath{\Lz^+_\cquark}}\xspace}

\def\Xib     {{\ensuremath{\Xires_\bquark}}\xspace}
\def\Xibz    {{\ensuremath{\Xires^0_\bquark}}\xspace}
\def\Xibm    {{\ensuremath{\Xires^-_\bquark}}\xspace}

\def\Omegab    {{\ensuremath{\Omegares^-_\bquark}}\xspace}

\def\to                 {\ensuremath{\rightarrow}\xspace}

\def\CP                {{\ensuremath{C\!P}}\xspace}

\def\AT#1     {\ensuremath{A_{\mathrm{T}}^{#1}}\xspace}           

\def\C#1      {\ensuremath{\mathcal{C}_{#1}}\xspace}                       
\def\Cp#1     {\ensuremath{\mathcal{C}_{#1}^{'}}\xspace}                    
\def\Ceff#1   {\ensuremath{\mathcal{C}_{#1}^{\mathrm{(eff)}}}\xspace}        
\def\Cpeff#1  {\ensuremath{\mathcal{C}_{#1}^{'\mathrm{(eff)}}}\xspace}       
\def\Ope#1    {\ensuremath{\mathcal{O}_{#1}}\xspace}                       
\def\Opep#1   {\ensuremath{\mathcal{O}_{#1}^{'}}\xspace}                    

\newcommand{\tev}{\ifthenelse{\boolean{inbibliography}}{\ensuremath{~T\kern -0.05em eV}\xspace}{\ensuremath{\mathrm{\,Te\kern -0.1em V}}}\xspace}
\newcommand{\gev}{\ensuremath{\mathrm{\,Ge\kern -0.1em V}}\xspace}
\newcommand{\mev}{\ensuremath{\mathrm{\,Me\kern -0.1em V}}\xspace}
\newcommand{\kev}{\ensuremath{\mathrm{\,ke\kern -0.1em V}}\xspace}
\newcommand{\ev}{\ensuremath{\mathrm{\,e\kern -0.1em V}}\xspace}
\newcommand{\gevc}{\ensuremath{{\mathrm{\,Ge\kern -0.1em V\!/}c}}\xspace}
\newcommand{\mevc}{\ensuremath{{\mathrm{\,Me\kern -0.1em V\!/}c}}\xspace}
\newcommand{\gevcc}{\ensuremath{{\mathrm{\,Ge\kern -0.1em V\!/}c^2}}\xspace}
\newcommand{\gevgevcccc}{\ensuremath{{\mathrm{\,Ge\kern -0.1em V^2\!/}c^4}}\xspace}
\newcommand{\mevcc}{\ensuremath{{\mathrm{\,Me\kern -0.1em V\!/}c^2}}\xspace}

\def\mub{\ensuremath{{\mathrm{ \,\upmu b}}}\xspace}

\def\gsim{{~\raise.15em\hbox{$>$}\kern-.85em
          \lower.35em\hbox{$\sim$}~}\xspace}
\def\lsim{{~\raise.15em\hbox{$<$}\kern-.85em
          \lower.35em\hbox{$\sim$}~}\xspace}

\def\pt         {\mbox{$p_{\mathrm{ T}}$}\xspace}

\def\tell1  {TELL1\xspace}
\def\ukl1   {UKL1\xspace}

\usepackage{cite} 
\usepackage{mciteplus}

\usepackage{longtable} 

\begin{document}

\renewcommand{\thefootnote}{\fnsymbol{footnote}}
\setcounter{footnote}{1}

\begin{titlepage}
\pagenumbering{roman}

\vspace*{-1.5cm}
\centerline{\large EUROPEAN ORGANIZATION FOR NUCLEAR RESEARCH (CERN)}
\vspace*{1.5cm}
\noindent
\begin{tabular*}{\linewidth}{lc@{\extracolsep{\fill}}r@{\extracolsep{0pt}}}
\ifthenelse{\boolean{pdflatex}}
{\vspace*{-2.7cm}\mbox{\!\!\!\includegraphics[width=.14\textwidth]{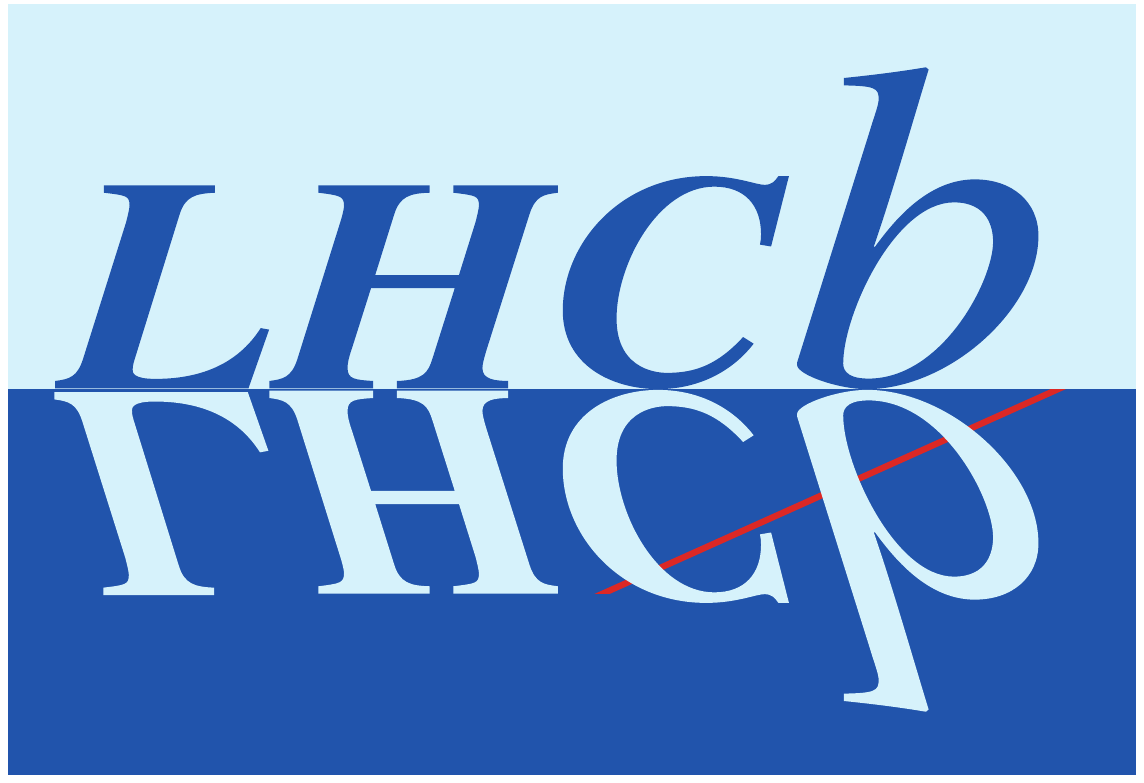}} & &}%
{\vspace*{-1.2cm}\mbox{\!\!\!\includegraphics[width=.12\textwidth]{lhcb-logo.eps}} & &}%
\\
 & & CERN-EP-2016-201 \\  
 & & LHCb-PAPER-2016-031 \\  
 & & January 4, 2017 \\ 
 & & \\
\end{tabular*}

\vspace*{4.0cm}

{\normalfont\bfseries\boldmath\huge
\begin{center}
  Measurement of  the $b$-quark production cross-section in 7 and 13~TeV $pp$ collisions 
  \end{center}
}

\vspace*{2.0cm}

\begin{center}
The LHCb collaboration\footnote{Authors are listed at the end of this paper.}
\end{center}

\vspace{\fill}

\begin{abstract}
  \noindent
Measurements of the cross-section for producing \bquark  quarks in the reaction $pp\to b\overline{b} X$ are reported in 7 and 13~TeV collisions at the LHC as a function of the pseudorapidity $\eta$ in the range $2<\eta<5$ covered by the acceptance of the LHCb experiment.  The measurements are done using semileptonic decays of $b$-flavored hadrons decaying into a ground-state charmed hadron in association with a muon. The cross-sections in the covered $\eta$ range are $72.0\pm 0.3\pm6.8~\mu$b and $144\pm 1\pm 21~\mub$ for 7 and 13~TeV. The ratio is $2.00\pm0.02\pm0.26$, where the quoted uncertainties are statistical and systematic, respectively.
The agreement with theoretical expectation is good at both 7~TeV, and 13~TeV.   
\end{abstract}

\vspace*{2.0cm}

\begin{center}
  Published in Phys.~Rev.~Lett. 
  \end{center}

\vspace{\fill}

{\footnotesize 
\centerline{\copyright~CERN on behalf of the \lhcb collaboration, licence \href{http://creativecommons.org/licenses/by/4.0/}{CC-BY-4.0}.}}
\vspace*{2mm}

\end{titlepage}


\renewcommand{\thefootnote}{\arabic{footnote}}
\setcounter{footnote}{0}



\pagestyle{plain} 
\setcounter{page}{1}
\pagenumbering{arabic}


%


Production of \bquark quarks in high energy $pp$ collisions at the LHC provides a sensitive test of models based on quantum chromodynamics \cite{Cacciari:2012ny,*Kniehl:2011bk}. Searches for physics beyond the Standard Model (SM) often rely on the ability to accurately predict the production rates of $b$ quarks that can form backgrounds in combination with other high energy processes \cite{Halkiadakis:2014qda}. In addition, knowledge of the $b$-quark yield is essential for calculating the sensitivity of experiments testing the SM by measuring $\CP$-violating and rare decay processes \cite{Bediaga:2012py}. 

We present here measurements of production cross-sections for the average of $b$-flavored and $\overline{b}$-flavored hadrons, denoted $pp\to H_b X$, where $X$ indicates additional particles, in $pp$ collisions recorded by LHCb at both 7 and 13~TeV center-of-mass energies, and their ratio.  These measurements are made as a function of the $H_b$ pseudorapidity  $\eta$ in the interval  $2< \eta < 5$, where $\eta=-{\rm ln}\left[\tan(\theta/2)\right]$,
and $\theta$ is the angle of the weakly decaying $b$ or $\overline{b}$-hadron with respect to the proton direction. We report results over the full range of  $b$-hadron transverse momentum, \pt.  The $H_b$ cross-section has been previously measured at LHCb in 7~TeV collisions using semileptonic decays to $D^0\mu^-X$ \cite{Aaij:2010gn} and $b\to \jpsi X$ decays \cite{Aaij:2011jh}. Previous determinations were made at the Tevatron collider in $\overline{p}p$ collisions near 2~TeV center-of-mass energy \cite{Abbott:1999se,*Aaltonen:2009xn,*Aaltonen:2007zza}. Other  LHC experiments have also measured $b$-quark production characteristics at 7~TeV \cite{Chatrchyan:2011vh,*Aaij:2012jd,*Chatrchyan:2012hw,*Chatrchyan:2012xg,*Abelev:2012gx,*Aad:2012jga,*ATLAS:2013cia,*Aaij:2013noa}, and 13~TeV \cite{Khachatryan:2016csy}.  The method presented in this Letter is more accurate because the normalization is based on well-measured semileptonic $\Bzb$ and $B^-$ branching fractions, and the equality of semileptonic widths for all $b$-hadrons, in contrast to inclusive \jpsi production which relies on the assumption that the $b$-hadron particle species are produced in the same proportions as at LEP \cite{PDG}, or those that just use one specific $b$-hadron, which needs the $b$-hadron fractions to extrapolate to the total.

 The production cross-section for a hadron $H_b$ that contains either a $b$ or $\overline{b}$ quark, but not both, is given by
\begin{align}
\label{eq:eq1}
\sigma(pp\to H_b X) = & \mspace{20mu} \frac{1}{2}\left[\sigma(B^0)+\sigma(\Bzb)\right]+\frac{1}{2}\left[\sigma(B^+)+\sigma(B^-)\right] \\
&+\frac{1}{2}\left[\sigma(B_s^0)+\sigma(\Bsb)\right]+\frac{1+\delta}{2}\left[\sigma(\Lb)+\sigma(\Lbbar)\right], \nonumber
\end{align}
where $\delta$ is a correction that accounts for $\Xib$ and $\Omegab$ baryons; we ignore $B_c$ mesons since their production level is estimated to be only 0.1\% of $b$ hadrons \cite{Aaij:2014zsa}.

Our estimate of $\delta$ is based on a paper by Voloshin \cite{Voloshin:2015xxa}, in which two useful relations are given
\begin{align}
\label{eq:volo}
\Gamma(\Xibm\to\PXi^- X \mu^-\overline{\nu})=&~\Gamma(\Lb\to\Lz X \mu^-\overline{\nu}),\\\nonumber
\text{and}\quad \frac{\sigma(\Xibm)}{\sigma(\Lb)}=& ~0.11\pm0.03\pm 0.03,
\end{align}
where the latter is determined from Tevatron data, and the second uncertainty is assigned from the allowable SU(3) symmetry breaking. The $b$-hadron fractions determined there \cite{PDG} agree with the ones measured by LHCb for other $b$-flavored hadrons \cite{Aaij:2011jp}.
Since the lifetimes of the \Lb and \Xibm are equal within their uncertainties \cite{PDG}, assuming that the two branching fractions are equal gives us an estimate of 0.11 for the $\PXi_b^-/\Lb$ semileptonic decay ratio. However, this must be doubled, using isospin invariance, to account for the $\Xibz$. To this we must add the \Omegab contribution, taken as 15\%  of the $\PXi_b$, thus arriving at an estimate of $\delta$ of $0.25\pm 0.10$, where the uncertainty is the one in Eq.~\ref{eq:volo} added in quadrature to our estimate of the uncertainties from assuming isospin and lifetime equalities.

To measure these cross-sections we determine the signal yields of $b$ decays into a charm hadron plus a muon for a given integrated luminosity ${\cal{L}}$ and correct for various efficiencies described below.  Explicitly
\begin{align}
\label{eq:sigmab}
\sigma(pp\to H_b X) = &  \frac{1}{2\cal{L}}\left\{\left[\frac{n(D^0\mu)}{\epsilon_{D^0} \times {\cal{B}}_{D^0}}
+ \frac{n(D^+\mu)}{\epsilon_{D^+} \times {\cal{B}}_{D^+}}\right]\frac{1}{{\cal{B}}(B\to D X \mu\nu)}\right.\\ \nonumber
& +\left[\frac{n(D_s^+\mu)}{\epsilon_{D_s^+} \times {\cal{B}}_{D_s^+}}\right]\frac{1}{{\cal{B}}(B_s\to D_s X\mu\nu)}
+ \left.\left[\frac{n(\Lc\mu)}{ \epsilon_{\Lc} \times {\cal{B}}_{\Lc}}\right]\frac{1+\delta}{{\cal{B}}(\Lb\to \Lc X\mu\nu)}\right\},
\end{align}
where $n(X_c\mu$) means the number of detected charm hadron plus muon events and their charge-conjugates, with corresponding efficiencies denoted by $\epsilon_{X_c}$. The charm branching fractions, $ {\cal{B}}_{X_c}$,  used in this analysis, along with their sources, are listed in the supplemental material. The PDG average is used for the $D^0$ and $D_s^+$ modes \cite{PDG}. For the $D^+$ mode there is only one measurement by CLEO III, so that is used \cite{Bonvicini:2013vxi}. For the \Lc we average measurements by BES III \cite{Ablikim:2015flg} and  Belle \cite{Zupanc:2013iki}. The expression ${\cal{B}}(B\to D X \mu\nu)$ denotes the average branching fraction for $\Bzb$ and $B^-$ semileptonic decays.

The \Bzb and \Bm semileptonic branching fractions are obtained with a somewhat different procedure than that adopted by the PDG, whose actual estimate is difficult to derive from the posted information. We take three measurements that are mostly model-independent and average them. The first one was made by CLEO using inclusive leptons at the $\PUpsilon(4S)$ resonance without distinguishing whether they are from \Bzb or \Bm meson decays \cite{Mahmood:2004kq}. The $\PUpsilon(4S)$, however, does not have an equal branching fraction into \Bzb\Bz and \Bm\Bp mesons. In fact the fraction into neutral $B$ pairs is $\alpha=0.486\pm 0.006$ \cite{PDG}, with the remainder going into charged $B$ pairs. Therefore to compute the \Bzb and \Bm semileptonic branching fractions we need to use the following coupled equations
\begin{align}
\label{eq:BslCLEO}
&\alpha{\cal{B}}_{\rm SL}^0+(1-\alpha){\cal{B}}_{\rm SL}^-=(10.91\pm 0.09\pm 0.24)\%, \\\nonumber
& {\cal{B}}_{\rm SL}^0/{\cal{B}}_{\rm SL}^-=\tau^0/\tau^-=0.927\pm 0.004,
\end{align}
where  $\tau^i$ are the lifetimes \cite{PDG}.
The numbers extracted from the solution are listed in Table~\ref{tab:BBSL}, along with direct measurements from CLEO \cite{Mahmood:2004kq}, BaBar \cite{Aubert:2006au}, and Belle \cite{Urquijo:2006wd}. These latter two analyses measure the semileptonic decays of $\Bzb$ and $B^-$ mesons separately. They do not cover the full momentum range so a correction has to be applied; this was done by the PDG \cite{PDG}. Since $D^0$ and $D^+$ mesons are produced in both $\Bzb$ and $B^-$ decays, we sum their yields and use the average semileptonic branching fraction for $\Bzb$ and $B^-$ decays, $\langle \Bzb +B^-\rangle$.

\begin{table}[b]
\begin{center}
\caption{Measured semileptonic decay branching fractions for \Bzb and \Bm mesons. Yhe correlation of the errors in the underlying measurements in the average is taken into account. The CLEO numbers result from solving Eq.~\ref{eq:BslCLEO}.}
\vskip 3mm
\label{tab:BBSL}
\def\arraystretch{1.1}
\begin{tabular}{ccl}
\hline
${\cal{B}}_{\rm SL}^0$  (\%) & ${\cal{B}}_{\rm SL}^-$  (\%)&  Source     \\
\hline
$10.49\pm 0.27$ &$11.31\pm 0.27$&CLEO\cite{Mahmood:2004kq}\\
$~9.64\pm 0.43$&$10.28\pm 0.47$&BaBar \cite{Aubert:2006au}\\
$10.46\pm 0.38$&$11.17\pm 0.38$  &Belle\cite{Urquijo:2006wd}\\\hline
$10.31\pm 0.19$&  $11.09\pm 0.20$  &Average \\\hline
\end{tabular}
\def\arraystretch{1.0}
\end{center}
\end{table}

The semileptonic $B$ branching fractions we use are listed in Table~\ref{tab:Bsemi}.  
Since we are detecting only $b\to c\mu\nu$ modes, we have to correct later for the fact that there is a small 1\% $b\to u\mu\nu$ component \cite{PDG}.
\begin{table}[t]
\begin{center}
\caption{Measured semileptonic decay branching fractions for $B$ mesons and  derived branching fractions for $\Bsb$ and \Lb based on the equality of semileptonic widths and the lifetime ratios.}
\vskip 3mm
\label{tab:Bsemi}
\def\arraystretch{1.1}
\begin{tabular}{lcccc}
\hline
Particle  &   $\tau$ (ps) & ${\cal{B}}_{\rm SL}$  (\%)& $\Gamma_{\rm SL}$  (ps$^{-1}$)& ${\cal{B}_{\rm SL}}$ (\%)  \\
                &  measured     & measured                        & measured                      &     to be used \\
\hline
$\Bzb$ & $1.519\pm 0.005$ & $10.31\pm 0.19$ &$0.0678\pm 0.0013$&  $10.31\pm 0.19$ \\
$B^-$ & $1.638\pm 0.004$ & $11.09\pm 0.20$ &$0.0680\pm 0.0013$&  $11.09\pm 0.20$ \\
$\langle \Bzb +B^-\rangle$&&$10.70\pm 0.19$ & & $10.70\pm 0.19$   \\
$\Bsb$ & $1.533\pm 0.018$ &  &&  $10.40\pm 0.30$ \\
$\Lb$ & $1.467\pm 0.010$ &  &&  $10.35\pm 0.28$ \\
\hline
\end{tabular}
\def\arraystretch{1.0}
\end{center}
\end{table}

The semileptonic widths $\Gamma_{\rm SL}$ are equal for all $H_b$ species used in this analysis except for a small correction for \Lb decays  (${\cal{B}}_{\rm SL}=\Gamma_{\rm SL}/\Gamma=\Gamma_{\rm SL}\times\tau$). This has proven to be true in the case of charm hadron decays even though the lifetimes of $D^0$ and $D^+$ differ by a factor of 2.5. The
decays of the \Lb are slightly different due to the absence of the chromomagnetic correction that affects $B$-meson decays but is absent in $b$ baryons~\cite{Manohar:1993qn,Bigi:1993fe,Bigi:2011gf}. Thus $\Gamma_{\rm SL}$, and also ${\cal{B}}_{\rm SL}$, are increased for the \Lb by ($4\pm$2)\% \cite{Aaij:2011jp}.

The input for the \Bsb lifetime listed in Table~\ref{tab:Bsemi} uses only measurements in the flavor-specific decay $\Bsb\to D_s^+\pi^-$ from CDF \cite{Aaltonen:2011qsa} and LHCb \cite{Aaij:2014sua}. Other measurements can in principle be used, e.g. in $\jpsi\phi$ or $\jpsi f_0(980)$ final states, but they then involve also determining $\Delta\Gamma_s$. Older measurements involving semileptonic decays are suspected of having larger uncontrolled systematic uncertainties \cite{Stone:2014pra}.  Finally, the \Lb lifetime is taken from the HFAG average \cite{HFAG}.

Corrections due to cross-feeds among the modes, for example from $\Bsb\to D K \mu^- X$ events or $\Lb\to\D N \mu^- X$ decays are well below our sensitivity, and thus we do not include them.

The data used here corresponds to integrated luminosities of 284.10$\pm$4.86~pb$^{-1}$ collected at 7~TeV and 4.60$\pm$0.18~pb$^{-1}$~at 13~TeV \cite{Aaij:2014ida} where special triggers were implemented to minimize uncertainties.   The \lhcb detector~\cite{Alves:2008zz,LHCb-DP-2014-002} is a single-arm forward
spectrometer covering the \mbox{pseudorapidity} range $2<\eta <5$. Components include a high-precision tracking system
consisting of a silicon-strip vertex detector surrounding the $pp$
interaction region, a large-area silicon-strip detector located
upstream of a dipole magnet with a bending power of about
$4{\rm\,Tm}$, and three stations of silicon-strip detectors and straw
drift tubes placed downstream of the magnet.
Different types of charged hadrons are distinguished using information
from two ring-imaging Cherenkov detectors (RICH).
Muons are identified by a
system composed of alternating layers of iron and multiwire
proportional chambers.

Events of potential interest are triggered by the identification of a muon in real time with a minimum \pt of 1.48~GeV in the 7~TeV data \cite{LHCb-DP-2012-004}, and 0.9~GeV in the 13~TeV data (further restricted in the higher level trigger to $\pt>1.3$~GeV) \cite{Aaij:2016rxn}. In addition, to test for inconsistency with production at the primary vertex (PV), the  $\chi^2_{\rm IP}$  for the muon is computed as the difference between the vertex fit $\chi^2$ of the PV reconstructed with and without the considered track. We require that  $\chi^2_{\rm IP}$  be larger  than 200 at 7~TeV (16 at 13~TeV), and in the 7~TeV data only, the impact parameter of the muon must be greater than 0.5~mm. There is a prescale by a factor of two for both energies and an additional prescale of a factor of two for the $D^0\mu^-$ channel in the 7~TeV data.

These events are subjected to further requirements in order to select those with a charmed hadron decay which forms a vertex with the identified muon that is detached from the PV.  The charmed hadron must not be consistent with originating from the PV.  We use the decays $D^0\to K^-\pi^+$, $D^+\to K^-\pi^+\pi^+$, $D_s^+\to K^+K^-\pi^+$, and $\Lc\to p K^-\pi^+$. (The related branching fractions are given in the supplemental material.)
The RICH system is used to determine a likelihood  for each particle hypothesis. We
use selections on the differences of  log-likelihoods (${\cal{L}}$)to separate protons from kaons and pions, ${\cal{L}}(p)-{\cal{L}}(K)>0$ and ${\cal{L}}(p)-{\cal{L}}(\pi)>10$, kaons from pions ${\cal{L}}(K)-{\cal{L}}(\pi)>4$, and pions from kaons ${\cal{L}}(K)-{\cal{L}}(\pi)<4$ for 7~TeV and $<10$ for 13~TeV. In addition, in order to suppress background the average \pt of the charm hadron daughters must be larger than 700~MeV for three-body and 600~MeV for two-body decays, and the invariant mass of the charm hadron plus muon must range from approximately 3~GeV to 5~GeV. Furthermore, the charm plus $\mu$ vertex must be within a radius less than 4.8~mm from the beam-line to remove contributions of secondary interactions in the detector material due to long-lived particles, and the charm hadron must decay downstream of this vertex.

Since detection efficiencies vary over the available phase space, we divide the data into two-dimensional intervals in \pt of the charm plus $\mu$ system, and $\eta$, where the latter is determined from the relative positions of the charm plus $\mu$ vertex and the PV. We fit the data for each charm plus $\mu$ combination in each interval simultaneously in invariant mass of the charm hadron and ln($I\!P$/mm) variables, where $I\!P$ is the measured impact parameter of the charmed hadron with respect to the PV in units of mm.

\begin{figure}[tb]
		\centering
\includegraphics[scale=0.38]{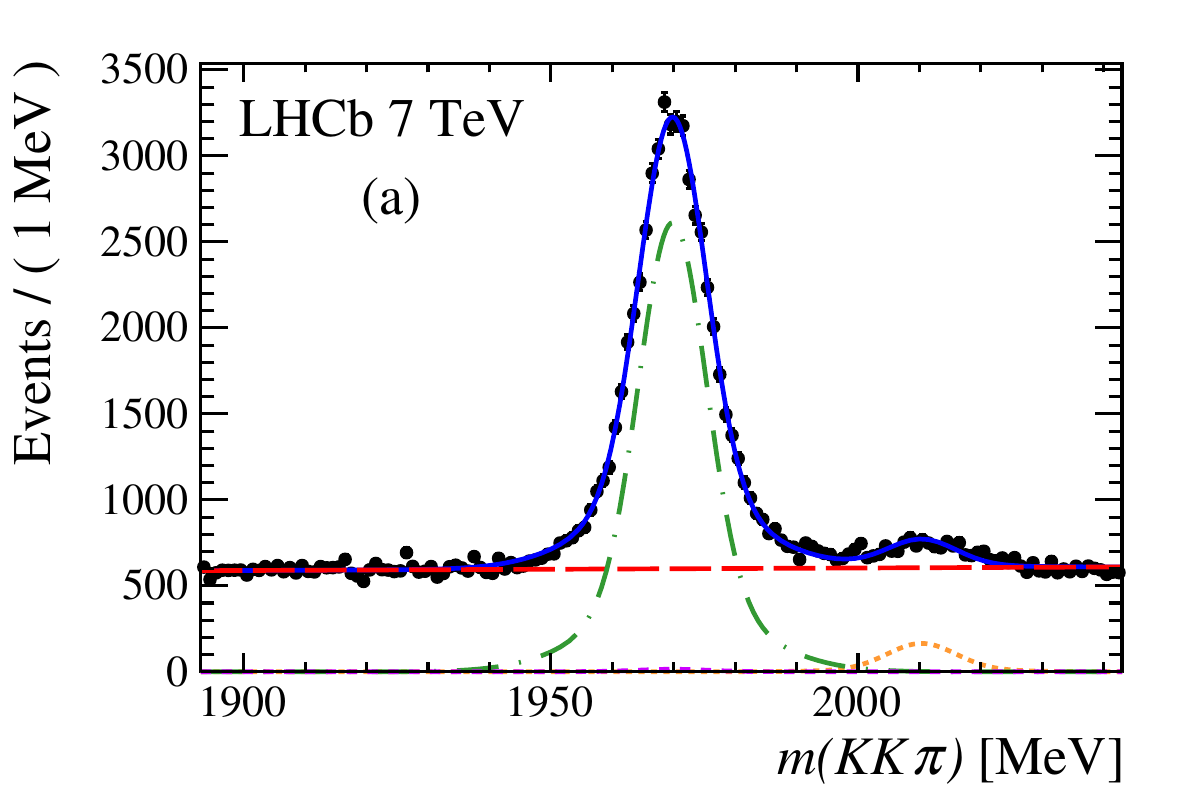}	\includegraphics[scale=0.38]{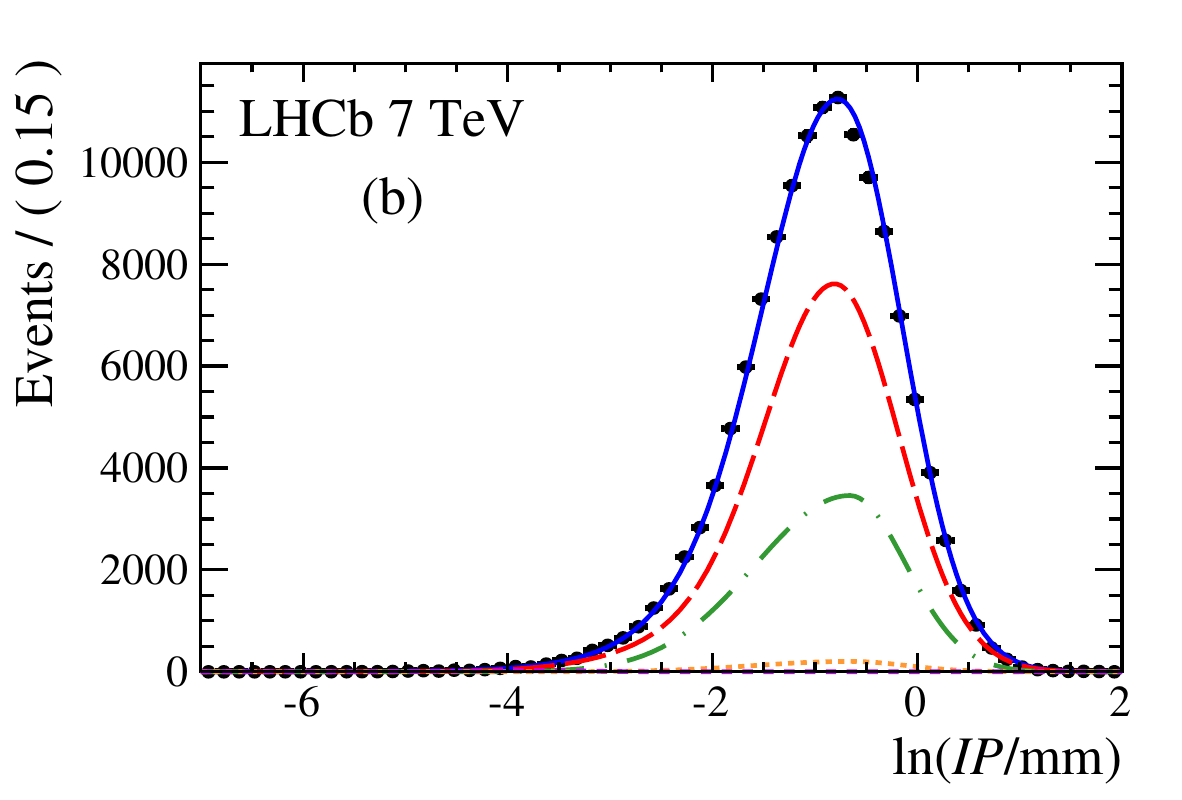}   
\includegraphics[scale=0.38]{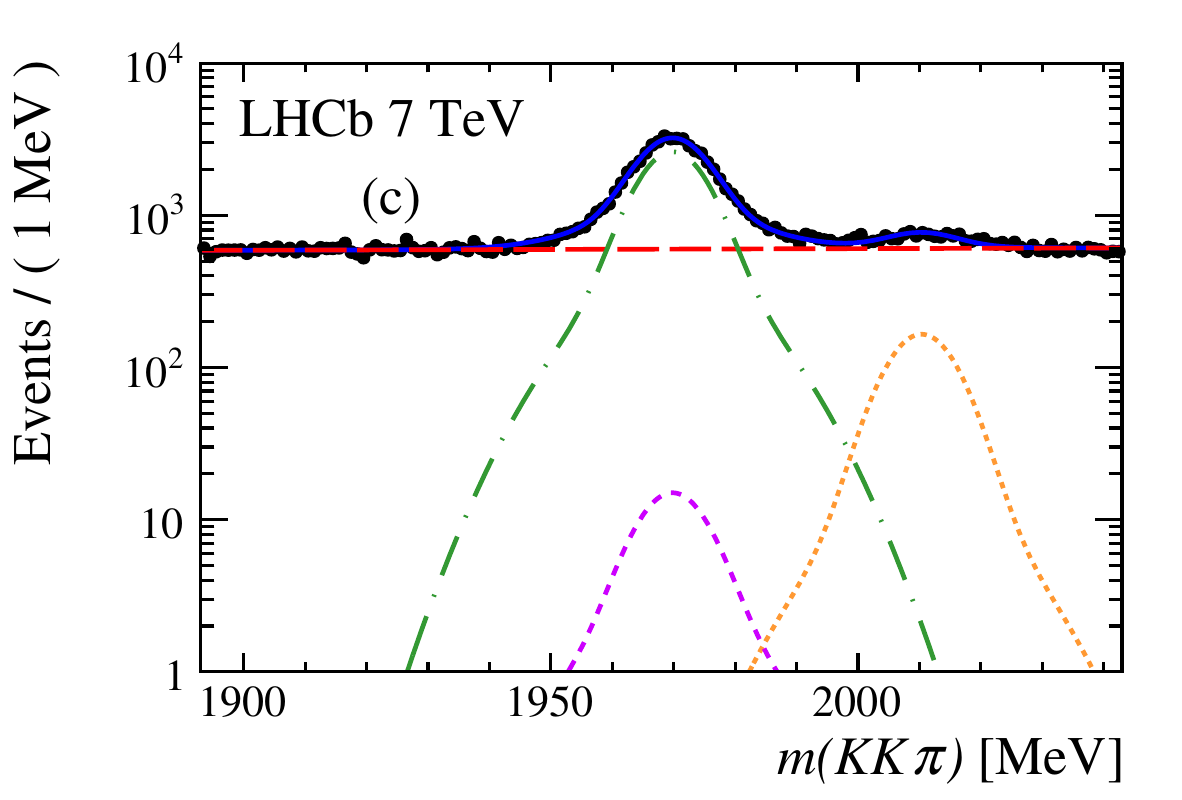}	\includegraphics[scale=0.38]{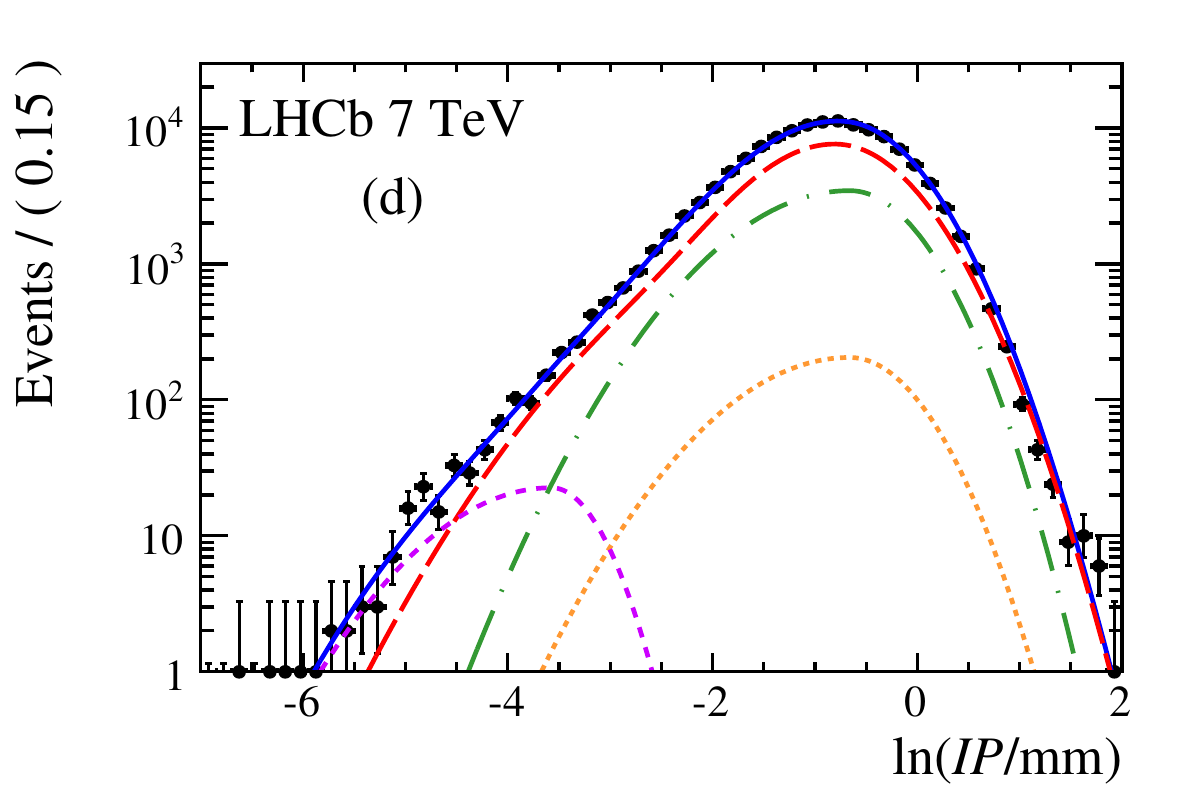}
		\caption{Fits to the $K^+K^-\pi^+$ invariant mass (a) and ln($I\!P$/mm) (b) distributions for data taken at  7~TeV data integrated over $2<\eta<5$. The data are shown as solid circles (black), and the overall fits as solid lines (blue). The dot-dashed (green) curve shows the $D_s^+$ signal from $b$ decay, while the dashed (purple) curve $D_s^+$ from prompt production. The dotted curve (orange) shows the $D^{*+}$ component. The dashed line (red) shows the combinatorial background. The same fits using a logarithmic scale are shown in (c) and (d). \label{fig:ds_example}}
\end{figure}

As an example of the fitting technique consider $D_s^+\mu^-$ candidates integrated over \pt and $\eta$ for the 7~TeV data.  Figure~\ref{fig:ds_example}(a) shows the $K^+K^-\pi^+$ invariant mass spectrum, while (b) shows the ln($I\!P$/mm) distribution. The invariant mass signal is fit for the $D_s^+$ yield with a double-Gaussian function where the means of the two Gaussians are constrained to be the same. The common mean and the widths are determined in the fit. (A second double-Gaussian shape is used to fit the higher mass decay of $D^{*+}\to \pi^+D^0, D^0\to K^+K^-$, an additional consideration only in this mode.) The ln($I\!P$/mm) shape of the signal component, determined by simulation,  is a bifurcated Gaussian where the peak position and width parameters are determined by the fit. The combinatorial background is modeled with a linear shape. (The other modes at both energies are shown in the supplemental material.)
The signal yields for charm hadron plus muon candidates integrated over $\eta$ are also given in the supplemental material.

The major components of the  total efficiency are the offline and trigger efficiencies.  The latter is measured with respect to the offline, which has several components from tracking, particle identification, event selection and overall event size cuts. These have been evaluated in a data-driven manner whenever possible. Only the event selection efficiencies have been simulated. Samples of simulated events, produced with the software described in Refs.~\cite{Sjostrand:2006za, Sjostrand:2007gs, LHCb-PROC-2010-056,*Lange:2001uf, *Golonka:2005pn, *Allison:2006ve, *Agostinelli:2002hh,*Clemencic:2011zza}, are used to characterize signal and background contributions. The particle identification efficiencies are determined from calibration samples of $D^{*+}\to\pi^+D^0$, $D^0\to K^-\pi^+$ decays for kaons and pions, and $\PLambda\to p\pi^-$ for protons. The trigger efficiencies including the muon identification efficiency are determined using samples of $b\to \jpsi X$, $\jpsi\to\mu^+\mu^-$ decays, where one muon is identified and the other used to measure the efficiencies.  For the overall sample they are typically 20\% for the 7~TeV data and 70\% for the 13~TeV data, only weakly dependent on $\eta$. The difference is caused primarily by the impact parameter cut  on the muon of 0.5~mm in the 7~TeV data.  The efficiency for the overall event size requirement is determined using $B^-\to\jpsi K^-$ decays where much looser criteria were applied. These efficiencies are all above 95\% and are determined with negligible uncertainties. The total efficiencies given as a function of $\eta$ and \pt for both energies are shown in the supplemental material.

There is dwindling efficiency toward small \pt values of the charmed hadron plus muon. Data in the regions with negligible efficiency  are excluded, and a correction is made using  simulation to calculate the fraction of events that fall within inefficient regions. These numbers are calculated for each bin of $\eta$ for 7~TeV and 13~TeV data separately, and the averages are 38\% at 7~TeV and 46\% at 13~TeV. The \pt distributions from simulation in each $\eta$ bin have been checked and found to agree within error with those observed in the data in bins with sufficient statistics.

The signal yields are obtained from fits that subtract the uncorrelated backgrounds. There are, however, two background sources that must be dealt with separately. One results from real charm hadron decays that form a vertex with a charged track that is misidentified as a muon and the other is from $b$ decays into two charmed hadrons where one decays either leptonically or semileptonically into a muon. In most cases the requirement that the muon forms a vertex with the charmed hadron eliminates this background, but some remains. 
 The background from fake muons combined with a real charmed hadron, and a real muon combined with a charm hadron from another $b$ decay as estimated from wrong-sign muon and hadron combinations is 0.7\% at 7~TeV and 2.0\% at 13~TeV.  The fake rates caused by $b$ decays to two charmed hadrons where one decays semileptonically have been evaluated from simulation and are about 2\%,
when averaged over all charmed species.

\begin{figure}[b]
\centering
\includegraphics[width=0.5\linewidth]{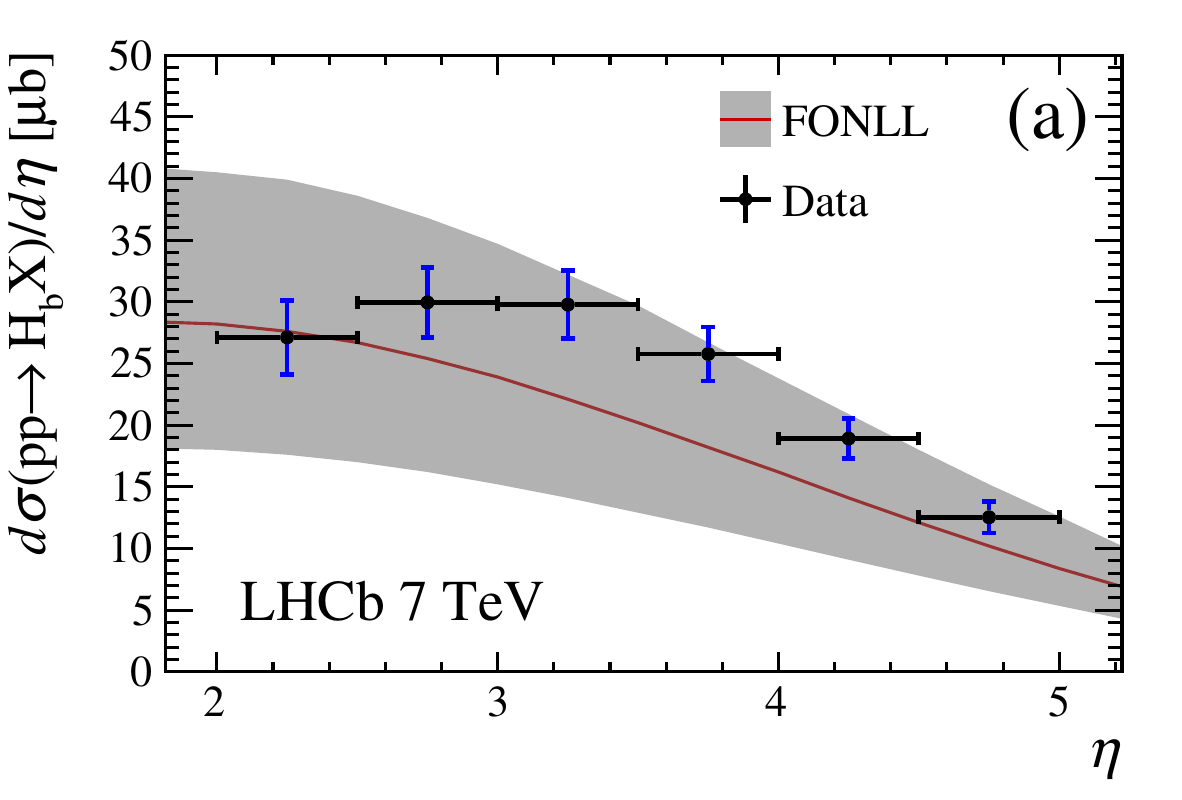}\includegraphics[width=0.5\linewidth]{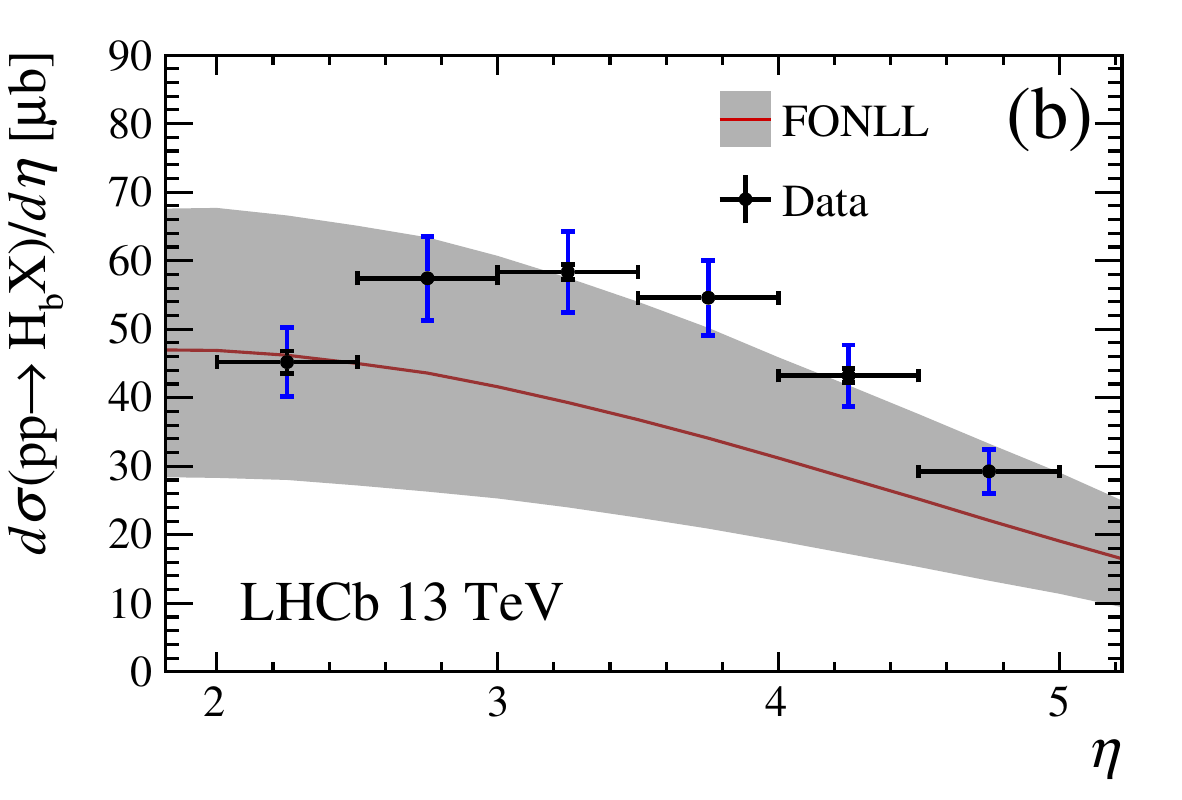}
\includegraphics[width=0.5\linewidth]{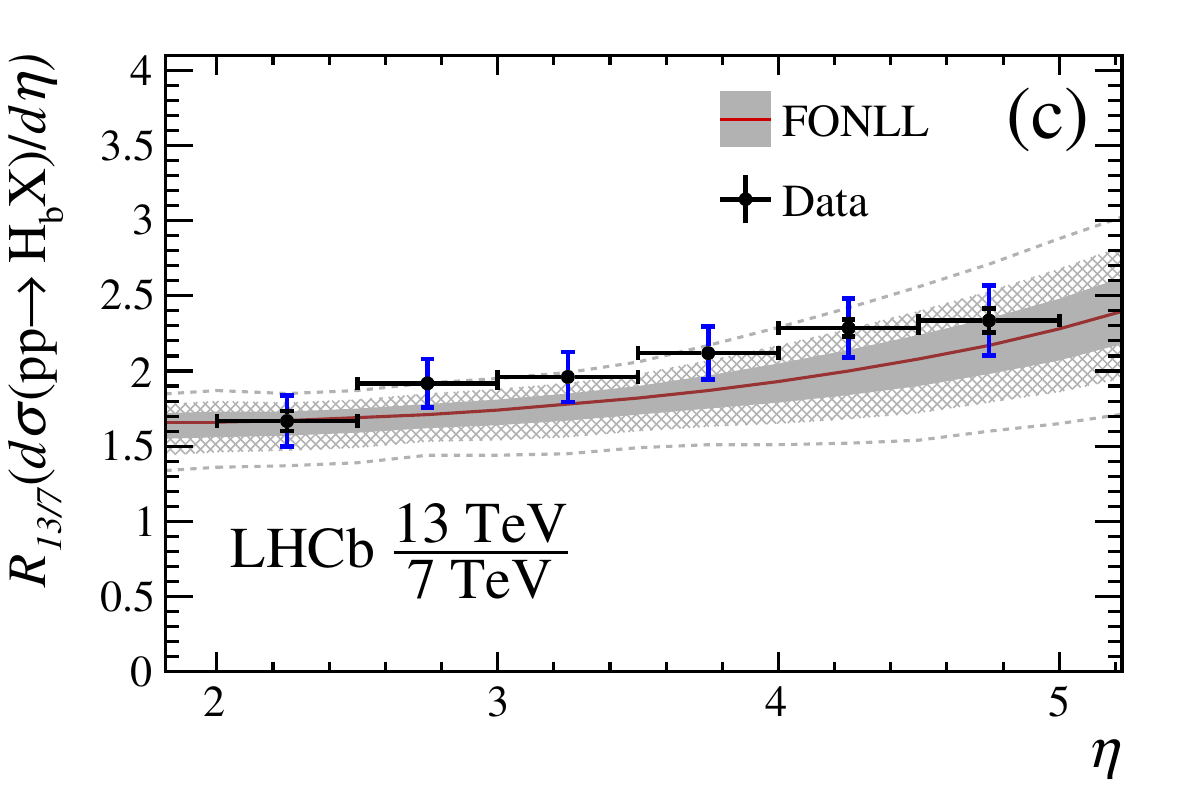}
\caption{\small The cross-section as a function of $\eta$ for $\sigma(pp\to H_b X)$, where $H_b$ is a hadron that contains either a $b$ or a $\overline{b}$ quark, but not both, at center-of-mass energies of 7~TeV (a) and 13~TeV (b). The ratio is shown in (c). The smaller error bars (black)  show the statistical uncertainties only, and the larger ones (blue)  have the systematic uncertainties added in quadrature. The solid line (red) gives the theoretical prediction, while the solid shaded band gives the estimated uncertainty on the predictions at $\pm1~\sigma$, the cross-hatched at $\pm2~\sigma$, and the dashes at $\pm3~\sigma$.}
\label{cross_section_7TeV_PtetaProjection}
\end{figure}

The inclusive $b$-hadron cross-sections as functions of $\eta$ are given in Fig.~\ref{cross_section_7TeV_PtetaProjection}, along with a theoretical prediction called FONLL \cite{Cacciari:2015fta}. These results are consistent with and supersede our previous results at 7~TeV  \cite{Aaij:2010gn}.
The ratio of cross-sections is predicted with less uncertainty, and indeed most of the experimental uncertainties (discussed below) also cancel, with the largest exception being the luminosity error.  In Fig.~\ref{cross_section_7TeV_PtetaProjection} (c), we compare the $\eta$-dependent cross-section ratio for 13~TeV divided by 7~TeV with the FONLL prediction.

The results as a function of $\eta$ are listed in Table~\ref{tab:bxresults}. 
The total cross-sections at 7 and 13~TeV integrated over $2<\eta<5$ are 
$72.0\pm 0.3\pm6.8~\mu$b and $144\pm 1\pm 21~\mu$b for 7 and 13~TeV. The ratio is $2.00\pm0.02\pm0.26$.
 This agrees with the theoretical prediction at 7~TeV of $62^{+28}_{-22}~\mu$b, and is a bit larger than the 13~TeV prediction of  $111^{+51}_{-44}~\mu$b. The measured ratio is consistent with the prediction of $1.79^{+0.21}_{-0.15}$.
\begin{table}[t]
\selectfont

\begin{center}
\caption{$pp\to H_b X$ cross-sections as a function of $\eta$ for 7~TeV and 13~TeV collisions and their ratio. The first uncertainty is
statistical and the second systematic. \label{tab:bxresults}}
\vskip 3mm
\label{tab:sys2}
\begin{tabular}{cccc}
\hline 
$\eta$ & 7~TeV ($\mub$) &  13~TeV ($\mub$) & Ratio 13/7\\\hline
2.0--2.5 &  13.6$\pm$0.2$\pm$1.5& 22.6$\pm$0.8$\pm$2.4& 1.67$\pm$0.07$\pm$0.16\\
2.5--3.0 &  15.0$\pm$0.1$\pm$1.4 & 28.7$\pm$0.4$\pm$3.0 & 1.92$\pm$0.03$\pm$0.16\\
3.0--3.5 & 14.9$\pm$0.1$\pm$1.4 & 29.2$\pm$0.6$\pm$2.9& 1.96$\pm$0.04$\pm$0.16\\
3.5--4.0 & 12.9$\pm$0.1$\pm$1.1 &  27.3$\pm$0.4$\pm$2.7 & 2.12$\pm$0.04$\pm$0.17\\
4.0--4.5 & \!~~9.5$\pm$0.1$\pm$0.8 &  21.6$\pm$0.5$\pm$2.2 & 2.29$\pm$0.06$\pm$0.19\\
4.5--5.0 &  \!~~6.3$\pm$0.1$\pm$0.6 & 14.6$\pm$0.5$\pm$1.5 &2.34$\pm$0.08$\pm$0.22\\

\hline
\end{tabular}
\end{center}
\end{table}

Systematic uncertainties are considerably larger than the statistical errors. The ones that are independent of $\eta$ are listed in Table~\ref{tab:sys1}.
The luminosity and muon trigger efficiency uncertainties in the ratio are each obtained by assuming a $-$50\% correlated error \cite{Aaij:2015rla}. The uncertainty in the tracking efficiency is given by taking 0.5\% per muon track and 1.5\% per hadron track \cite{Aaij:2014pwa}.  The various final states used to simulate the efficiencies can contribute to an overall efficiency change. This is estimated by taking the difference between the efficiencies of the higher multiplicity  $D^*\mu^-\nu$ states and $D^{**}\mu^-\nu$ states, where $D^{**}$ refers to excited states that decay into a charmed particle and pions, and taking into account the uncertainties on the measured branching fractions. These are then added in quadrature and referred to as the $b$ decay cocktail in Table~\ref{tab:sys1}. 

The  fraction of higher mass $b$-baryon states with respect to the \Lb  is given by $\delta=0.25\pm 0.10$, which represents a  40\% relative uncertainty that affects only the baryon contribution to Eq.~\ref{eq:sigmab}.

\begin{table}[b]
\begin{center}
\caption{\small Systematic uncertainties independent of $\eta$ on the $pp\to H_bX$ cross-sections at 7 and 13~TeV and their ratio.\label{tab:sys1}}
\vskip 3mm
\begin{tabular}{lccc}
\hline
Source & 7~TeV & 13 TeV & Ratio 13/7 \\
\hline
Luminosity &1.7\%  & 3.9\% &3.8\%\\
Tracking efficiency &  3.8\% & 6.3\%  &6.3\% \\
$b$ semileptonic ${\cal{B}}$ & 2.1\% & 2.1\%  & 0\\
Charm hadron ${\cal{B}}$ & 2.6\% & 2.6\% &0 \\
$b$ decay cocktail & 1.0\% & 1.0\% &0\\
Ignoring $b$ cross-feeds & 1.0\% & 1.0\% & 0\\
Background & 0.2\% & 0.3\% & 0 \\
$b\to u$ decays & 0.3\% & 0.3\% & 0\\
$\delta$ & 2.0\% & 2.0\% & 0.2\% \\
\hline
Total & 5.9\%&8.9\% &7.4\%\\
\hline
\end{tabular}
\end{center}
\end{table}

There are also $\eta$-dependent systematic uncertainties in the cross-section that arise from the trigger efficiency, the event selection, the hadron identification and the corrections for the low \pt region with low efficiencies. When added in quadrature with the $\eta$-independent uncertainties, the total errors range from (8.5--11.0)\% at 7~TeV to (8.7--9.7)\% at 13~TeV. There is some cancellation in the ratio giving a range of (5.6--7.3)\%.

In conclusion, new results for the $b\overline{b}$ production cross-section at 7~TeV are in good agreement with the original $\eta$-dependent cross-section measurement previously reported\cite{Aaij:2010gn}, and are in agreement with the theoretical prediction (FONLL) \cite{Cacciari:2015fta}. The 13~TeV results are slightly higher in magnitude than the theory,  and  generally agree with the shape and magnitude measured using inclusive $b\to\jpsi X$ decays \cite{Aaij:2015rla}. The cross-section ratio of 13 TeV to 7 TeV is $2.00\pm0.02\pm0.26$ and agrees with the theoretical prediction of $111^{+0.21}_{-0.15}~\!\mu b$.   Using multiplicative factors derived from PYTHIA 8 simulations of 4.1 at 7~TeV and 3.9 at 13~TeV \cite{Sjostrand:2007gs,LHCb-PROC-2010-056} we extrapolate to $b\overline{b}$ cross-sections over the full $\eta$ range of $\approx\! 295~\mu$b at 7~TeV and  $\approx\! 560~\mu$b at 13~TeV. 

\section*{Acknowledgements}

\noindent We express our gratitude to our colleagues in the CERN
accelerator departments for the excellent performance of the LHC. We
thank the technical and administrative staff at the LHCb
institutes. We acknowledge support from CERN and from the national
agencies: CAPES, CNPq, FAPERJ and FINEP (Brazil); NSFC (China);
CNRS/IN2P3 (France); BMBF, DFG and MPG (Germany); INFN (Italy);
FOM and NWO (The Netherlands); MNiSW and NCN (Poland); MEN/IFA (Romania);
MinES and FASO (Russia); MinECo (Spain); SNSF and SER (Switzerland);
NASU (Ukraine); STFC (United Kingdom); NSF (USA).
We acknowledge the computing resources that are provided by CERN, IN2P3 (France), KIT and DESY (Germany), INFN (Italy), SURF (The Netherlands), PIC (Spain), GridPP (United Kingdom), RRCKI and Yandex LLC (Russia), CSCS (Switzerland), IFIN-HH (Romania), CBPF (Brazil), PL-GRID (Poland) and OSC (USA). We are indebted to the communities behind the multiple open
source software packages on which we depend.
Individual groups or members have received support from AvH Foundation (Germany),
EPLANET, Marie Sk\l{}odowska-Curie Actions and ERC (European Union),
Conseil G\'{e}n\'{e}ral de Haute-Savoie, Labex ENIGMASS and OCEVU,
R\'{e}gion Auvergne (France), RFBR and Yandex LLC (Russia), GVA, XuntaGal and GENCAT (Spain), Herchel Smith Fund, The Royal Society, Royal Commission for the Exhibition of 1851 and the Leverhulme Trust (United Kingdom).

\clearpage
\newpage



\clearpage

\newpage
\ifx\mcitethebibliography\mciteundefinedmacro
\PackageError{LHCb.bst}{mciteplus.sty has not been loaded}
{This bibstyle requires the use of the mciteplus package.}\fi
\providecommand{\href}[2]{#2}

\newpage
\section*{Supplemental material}
The charm hadron branching fractions are given in Table~\ref{tab:results}. For the values taken from the PDG we use the ``Our Average" values in the listings rather than the values from 

\begin{table}[h]
\begin{center}
\caption{Charm hadron branching fractions for the decay modes used in this analysis.}
\vskip 3mm
\label{tab:results}
\begin{tabular}{lcl}
\hline
Decay &  ${\cal{B}}$ (\%) & Source\\
\hline
$D^0\to K^-\pi^+$ & $3.91\pm 0.05$ & PDG  \cite{PDG} \\\hline
$D^+\to K^-\pi^+\pi^+$ & $9.22\pm 0.17$ & CLEO III \cite{Bonvicini:2013vxi} \\\hline
$D_s^+\to K^-K^+\pi^+$ & $5.44\pm 0.18$ & PDG  \cite{PDG} \\\hline
$\Lc\to pK^-\pi^+$   &    $5.84 \pm 0.27 \pm 0.23$  &BES III\cite{Ablikim:2015flg} \\
                                   &     $6.84\pm0.24^{+0.21}_{-0.27}$ & Belle \cite{Zupanc:2013iki}\\
                                   &     $6.36\pm 0.35$ & Average$^\dagger$\\
\hline
\end{tabular}
\end{center}
\center{$^\dagger$Uncertainty increased due to discrepant central values from $\pm$0.25.}
\end{table}

The signal yields for charm hadron plus muon candidates integrated over $\eta$ are listed in Table~\ref{tab:integratedrates}.

\begin{table}[h]
\begin{center}
\caption{Signal yields of charm hadron plus $\mu$ events. Note that the 7~TeV $D^0$ result is prescaled in the trigger by a factor of two with respect to the others.}
\vskip 3mm
\label{tab:integratedrates}
\begin{tabular}{lcccc}
\hline
  & $D^0$ &$D^+$&$D_s^+$&$\Lc$ \\
\hline
~~\!7~TeV&  205\! 677$\pm$486   &   161\! 264$\pm$516 & 42\! 661$\pm$381   & 57\! 714$\pm$319 \\
13~TeV& ~56\! 560$\pm$277&  ~22\! 196$\pm$200  & ~\!5 \!013$\pm$91   &  ~\!6 \!442$\pm$94  \\
\hline
\end{tabular}
\end{center}
\end{table}

\begin{figure}[htb]
\centering
\includegraphics[scale=.3]{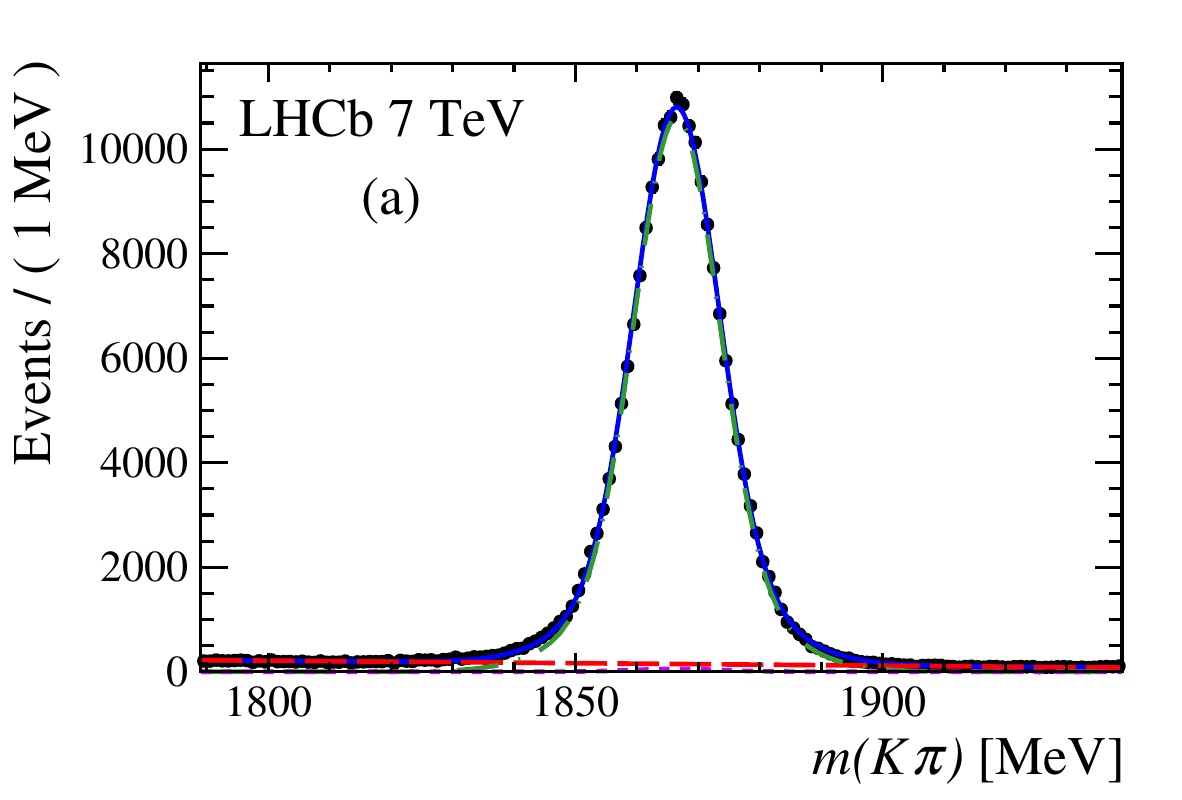}\includegraphics[scale=.3]{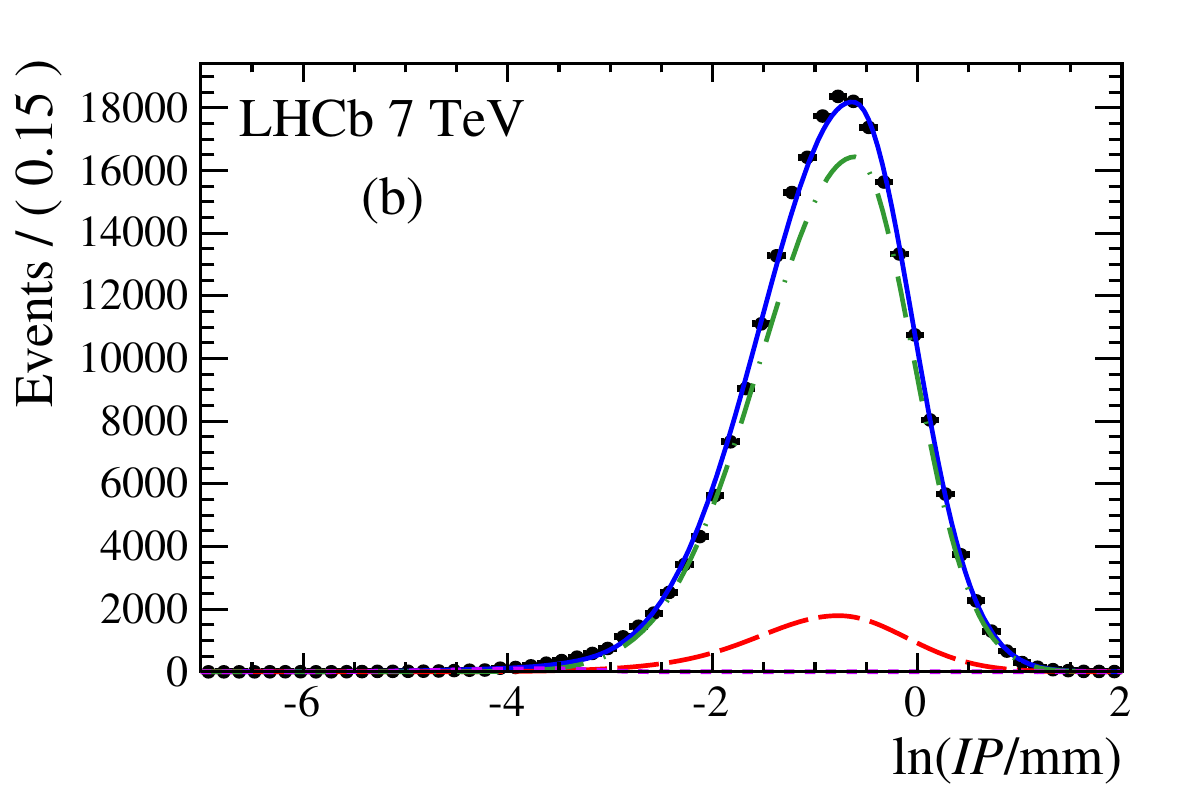}
\includegraphics[scale=.3]{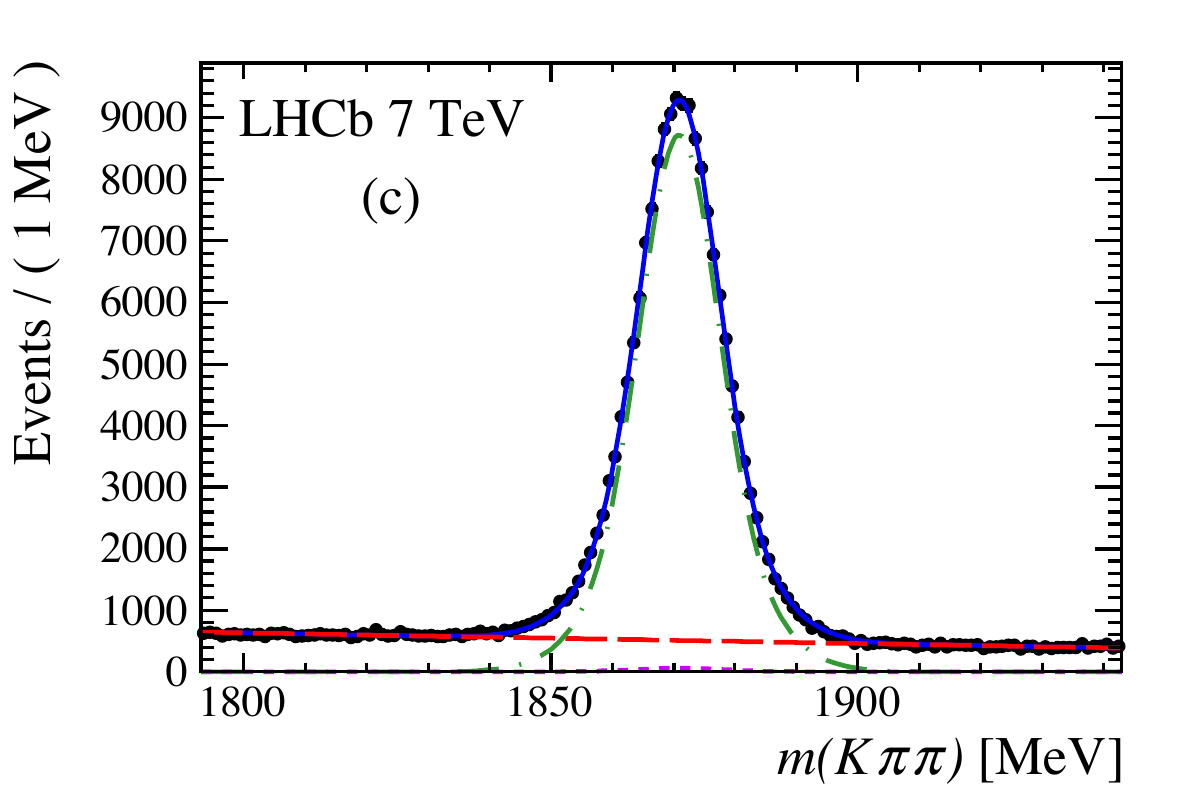}\includegraphics[scale=.3]{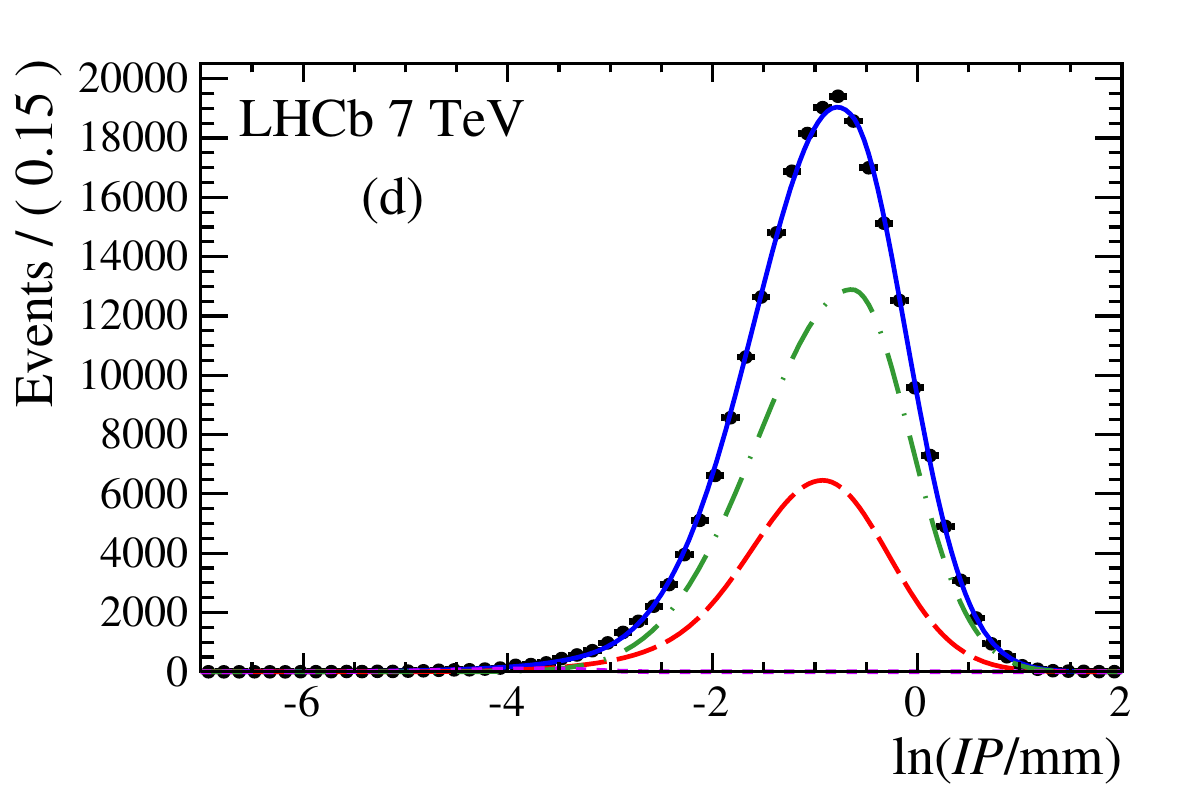} 	
\includegraphics[scale=.3]{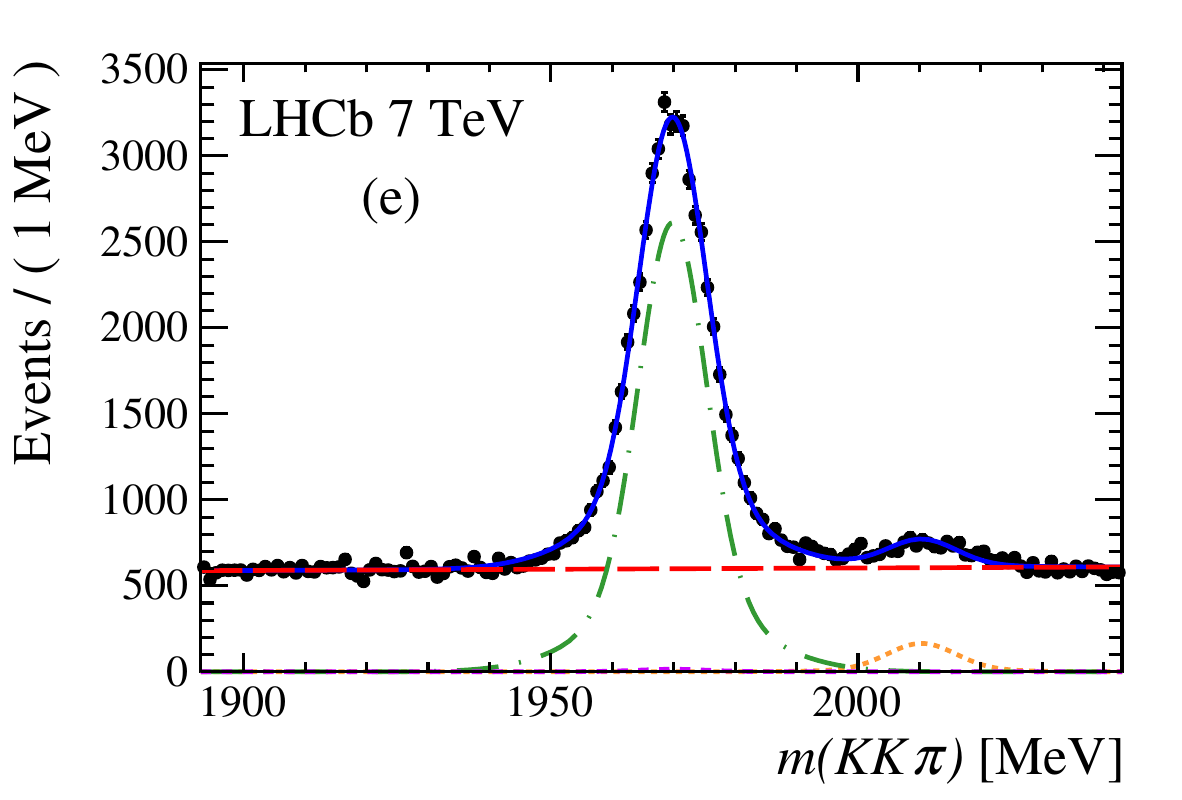}\includegraphics[scale=.3]{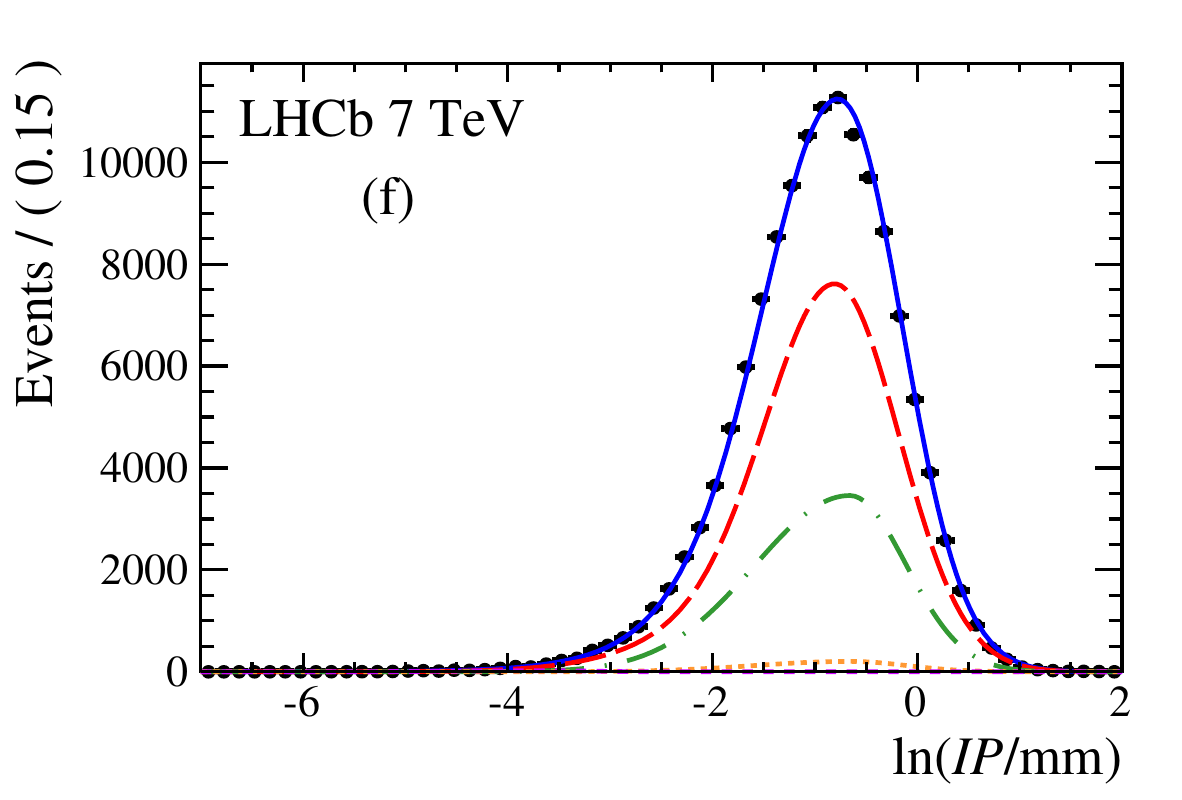}
\includegraphics[scale=.3]{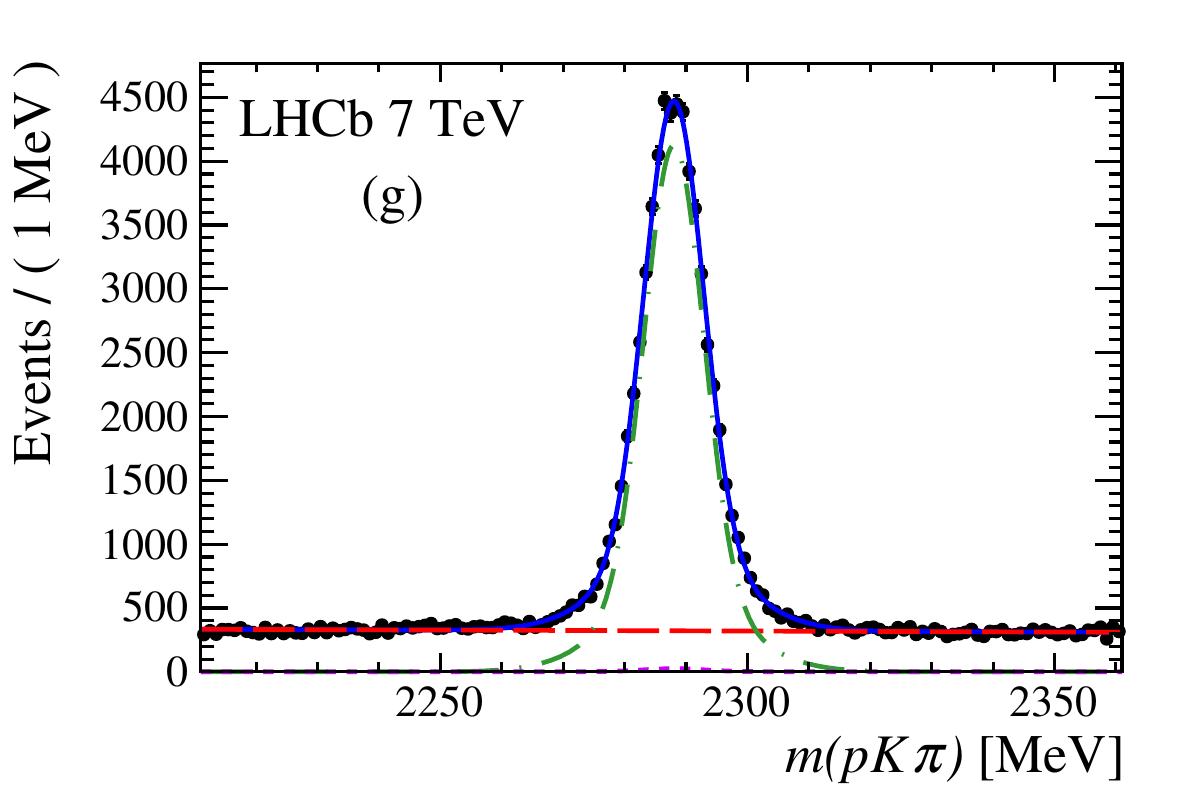}\includegraphics[scale=.3]{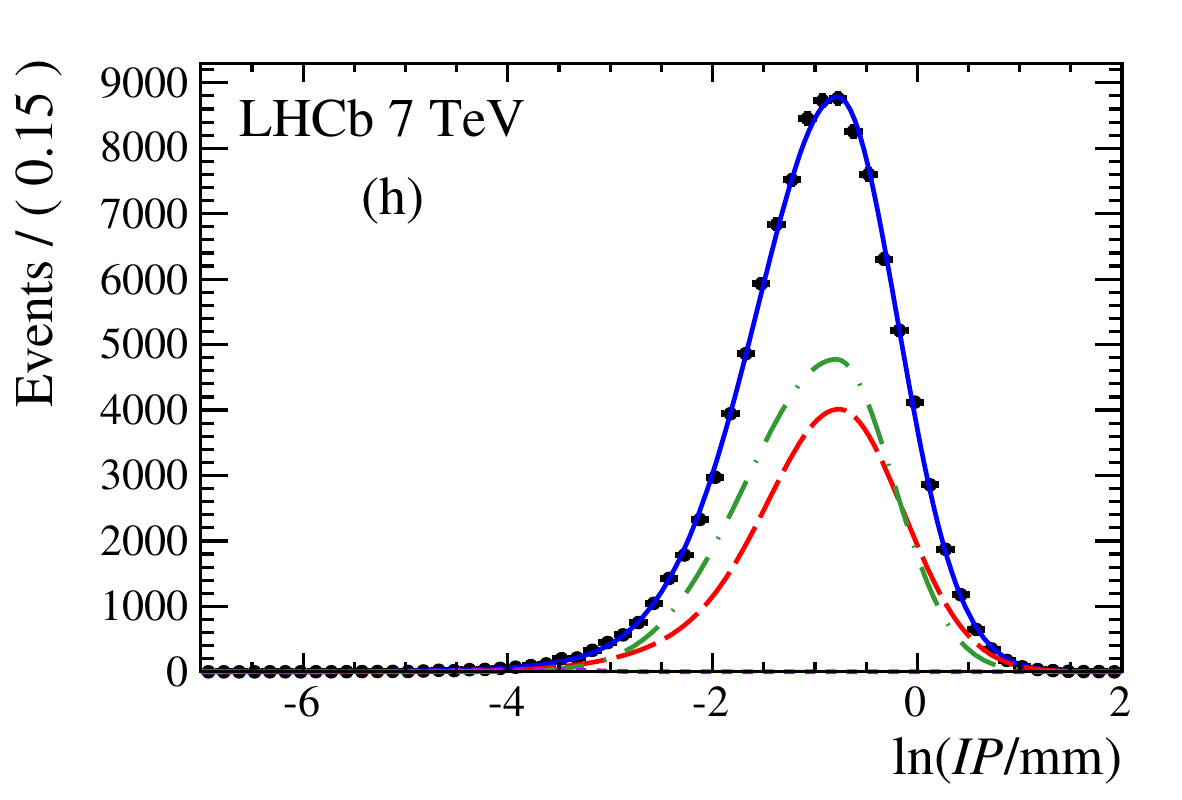}
\caption{Fits to the invariant masses and ln($I\!P$/mm) distributions integrated over $2<\eta<5$  for 7~TeV running. The data are shown as solid circles (black), and the overall fits as solid lines (blue). The dot-dashed (green) curves show the charm signals from $b$ decay, while the dashed (purple) curves  charm background from prompt production. The dashed line (red) shows the combinatorial background. The dotted curve (orange) shows the $D^{*+}$ component only for the $K^+K^-\pi^+$ mass distribution. (a) and (b) show $K^-\pi^+$ combinations,  (c) and (d) show $K^-\pi^+\pi^+$ combinations, (e) and (f) show $K^-K^+\pi^+$ combinations, and (g) and (h) show $pK^-\pi^+$ combinations.
 The fitting procedure is described in the text. \label{fig:d0_example7}}
\end{figure}

\begin{figure}[htb]
\centering
\includegraphics[scale=.3]{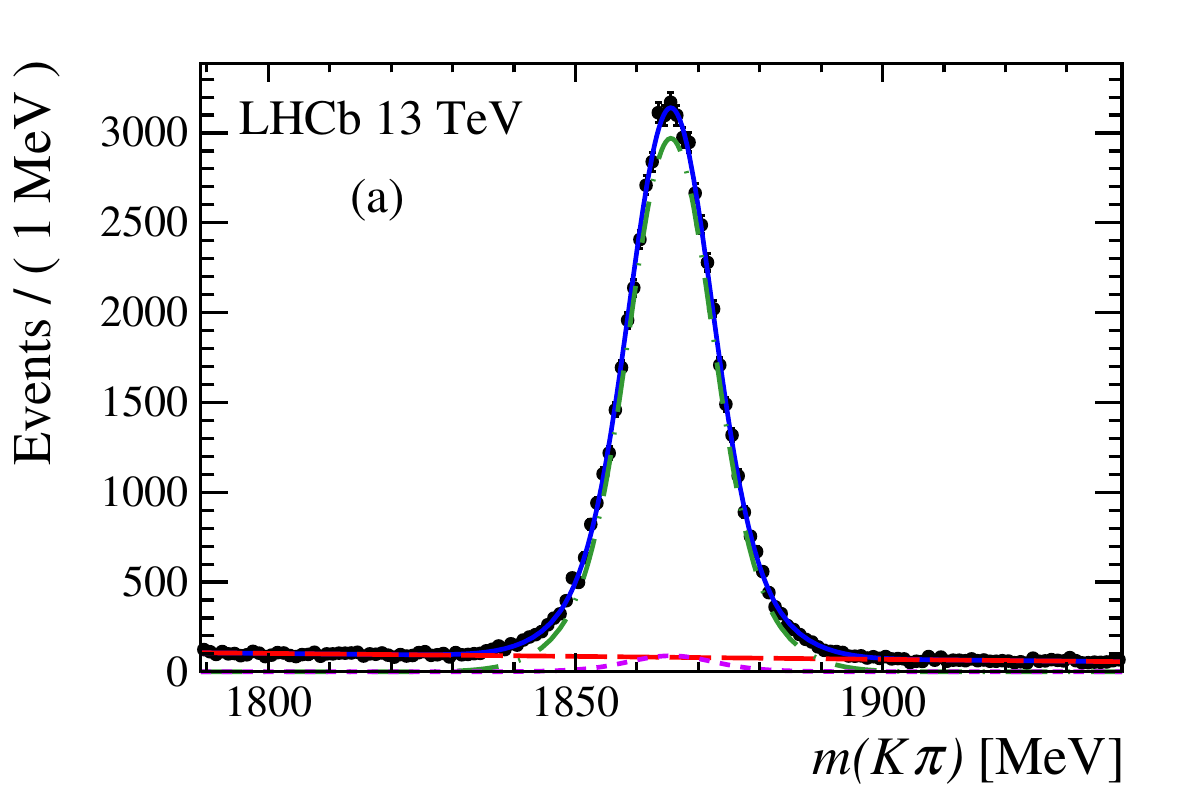}\includegraphics[scale=.3]{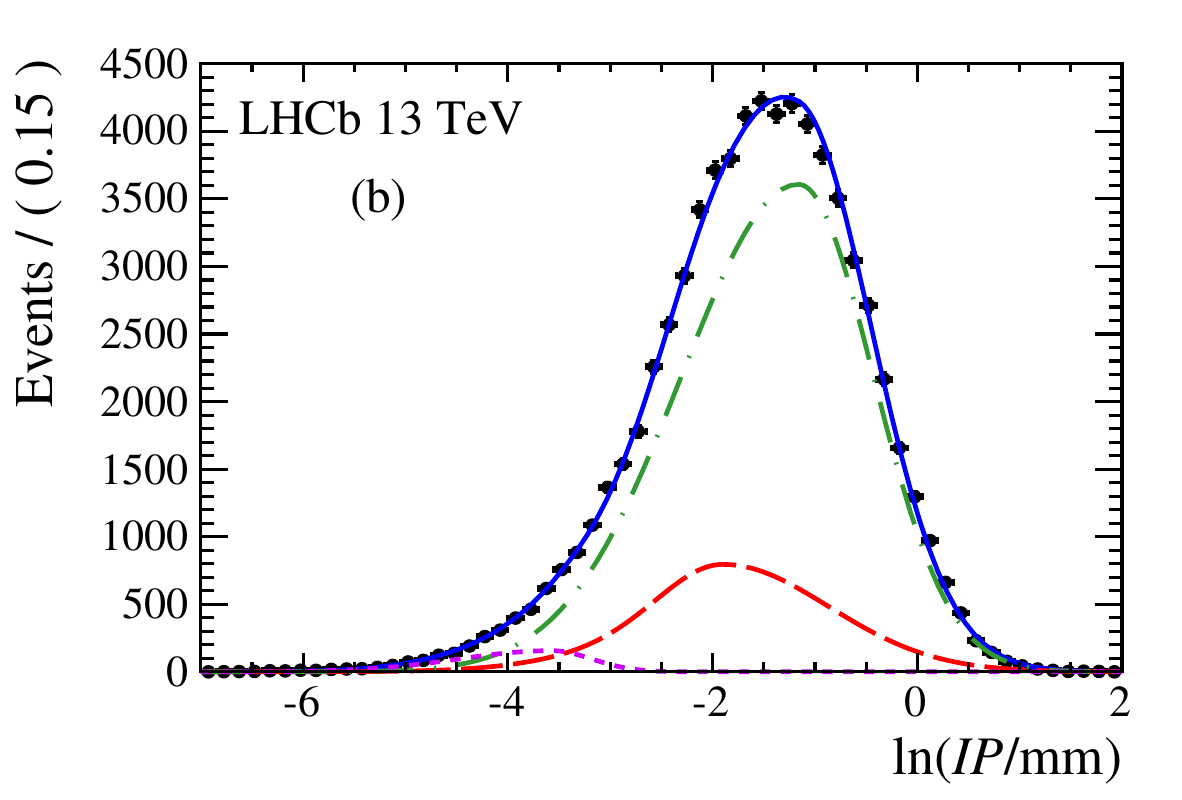}
\includegraphics[scale=.3]{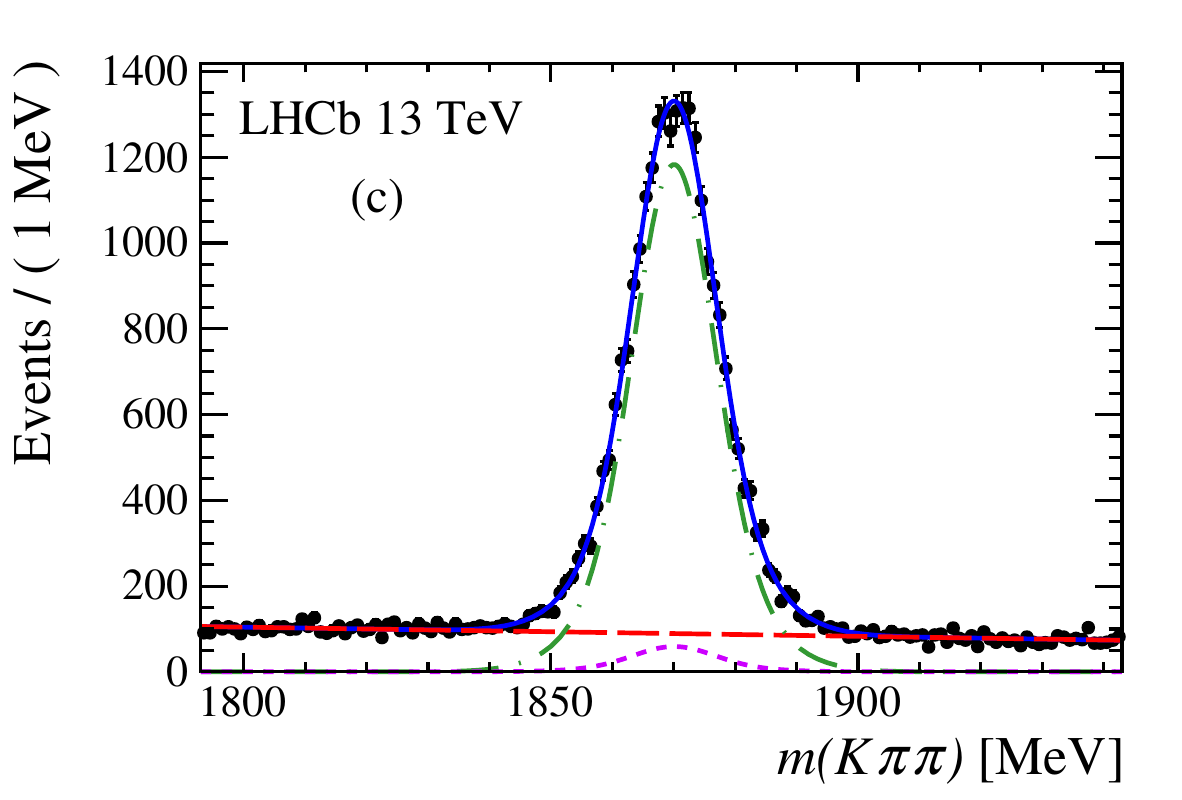}\includegraphics[scale=.3]{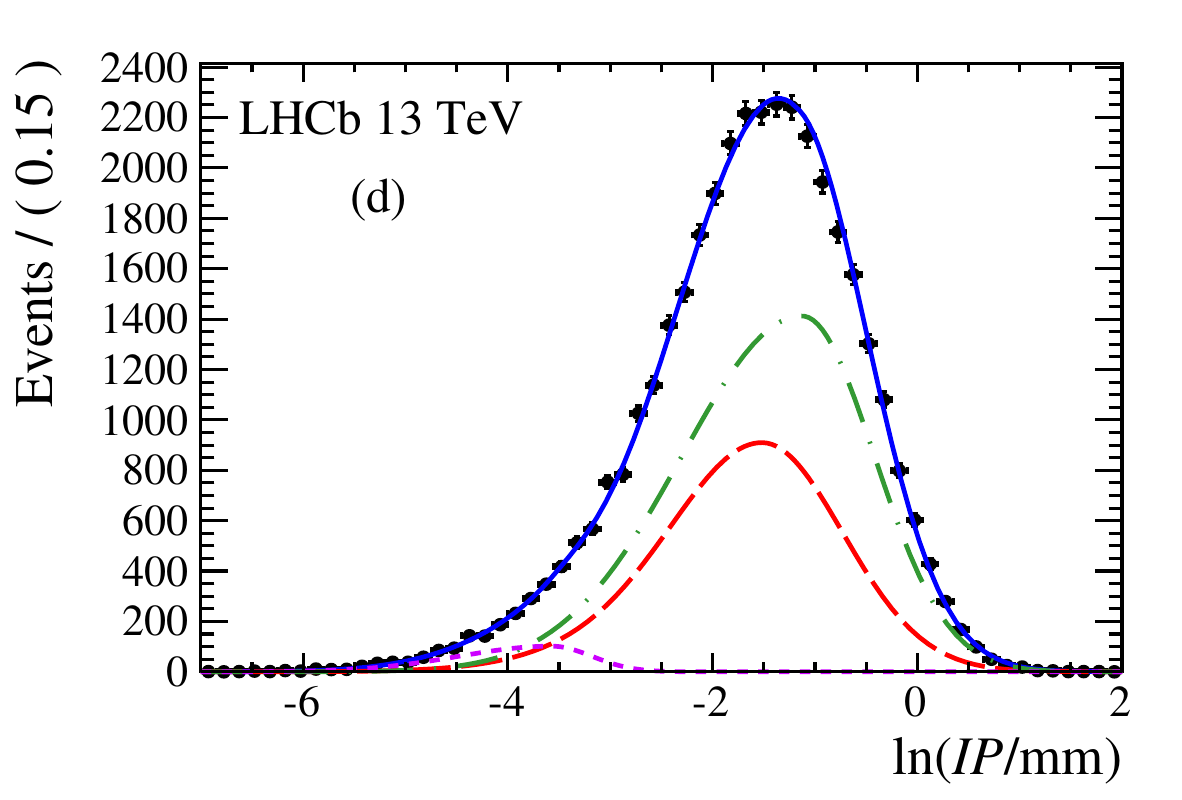} 	
\includegraphics[scale=.3]{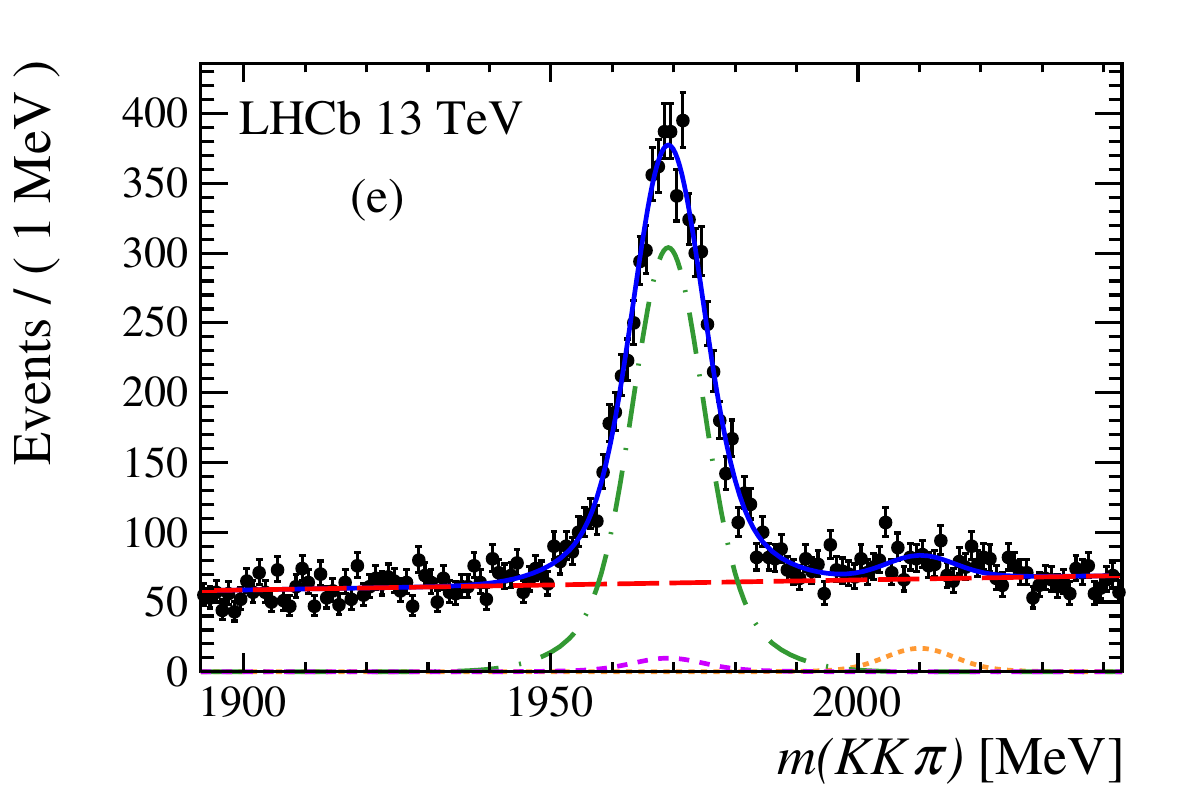}\includegraphics[scale=.3]{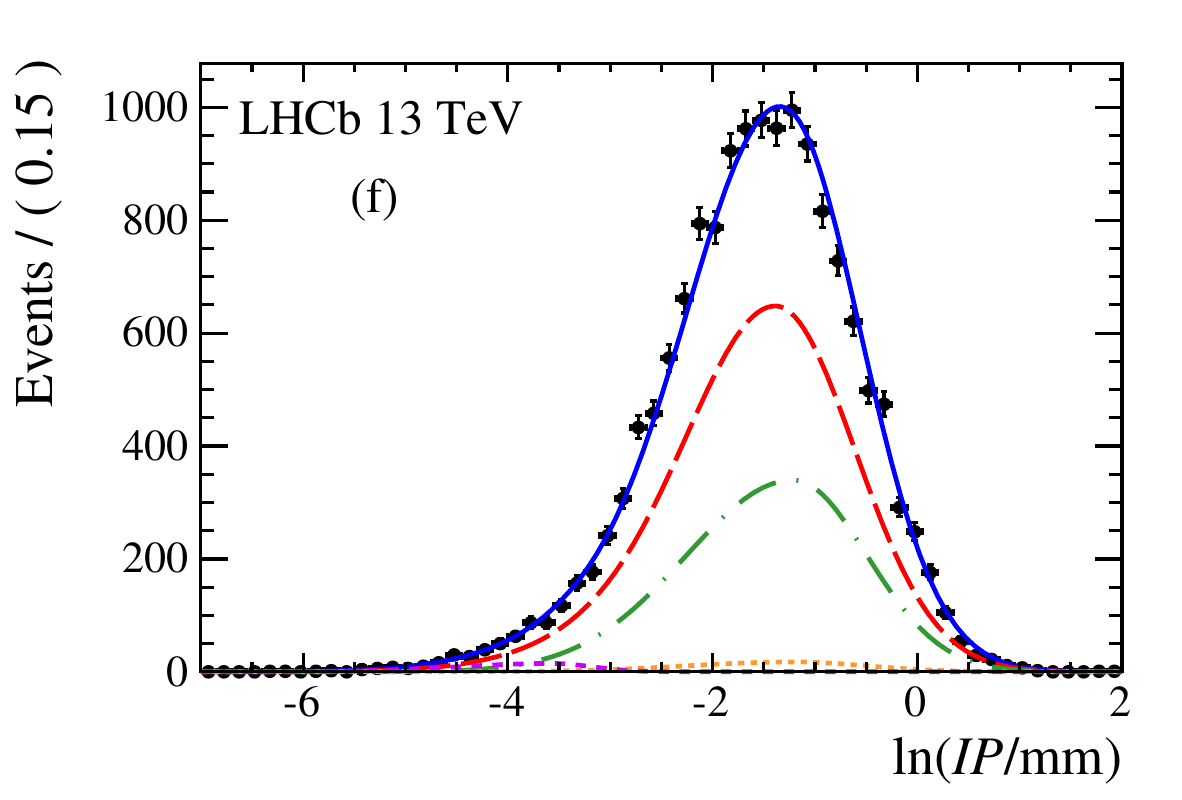}
\includegraphics[scale=.3]{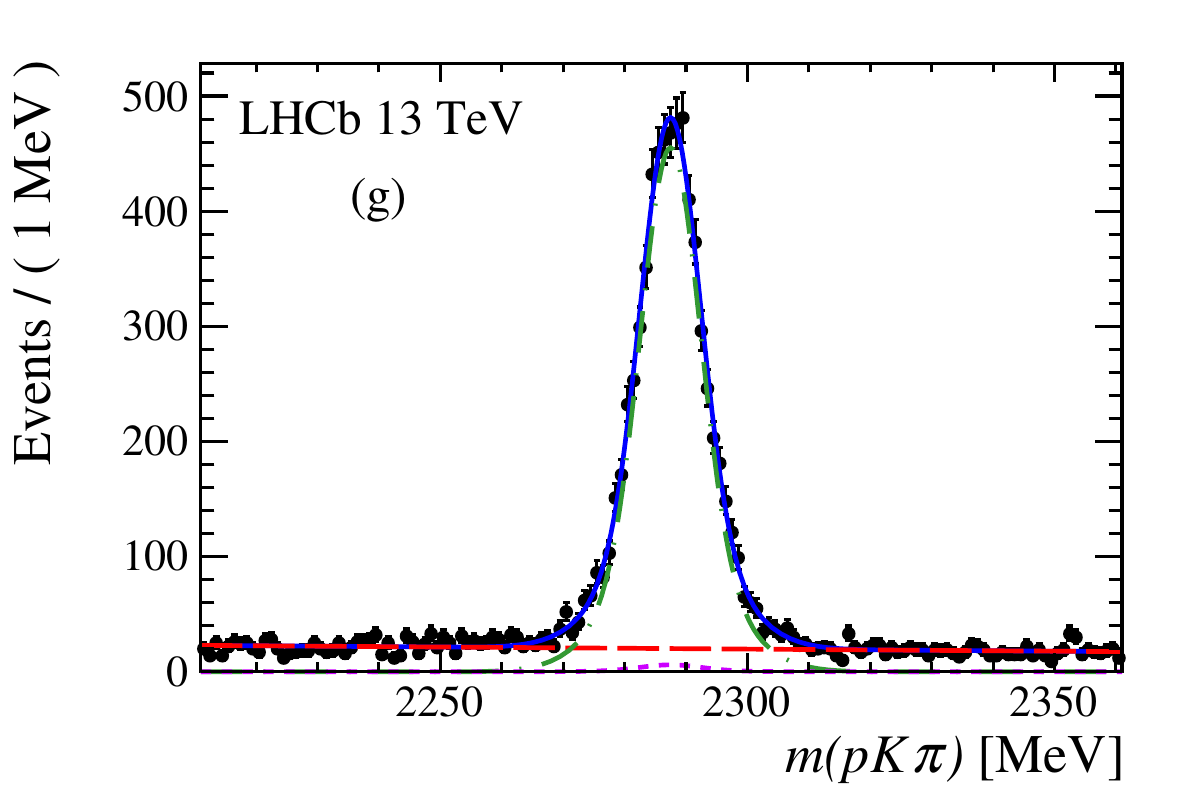}\includegraphics[scale=.3]{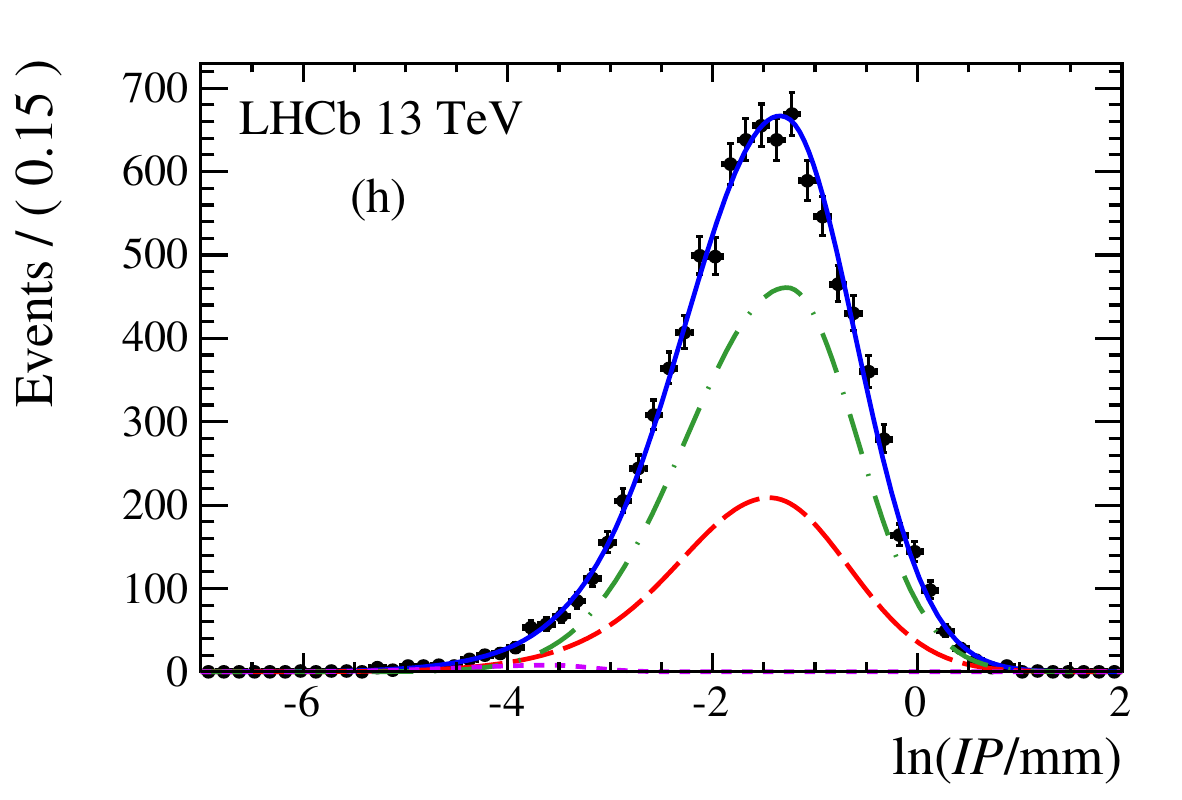}
\caption{Fits to the invariant masses and ln($I\!P$/mm) distributions integrated over  $2<\eta<5$  for 13~TeV running. The data are shown as solid circles (black), and the overall fits as solid lines (blue). The dot-dashed (green) curves show the charm signals from $b$ decay, while the dashed (purple) curves charm background from prompt production. The dashed lines (red) show the combinatorial background. The dotted curve (orange) shows the $D^{*+}$ component only for the $K^+K^-\pi^+$ mass distribution. (a) and (b) show $K^-\pi^+$ combinations,  (c) and (d) show $K^-\pi^+\pi^+$ combinations, (e) and (f) show $K^-K^+\pi^+$ combinations, and (g) and (h) show $pK^-\pi^+$ combinations.
 The fitting procedure is described in the text.  \label{fig:d0_example13}}
\end{figure}

\begin{figure}[tb]
\centering
\includegraphics[scale=0.8]{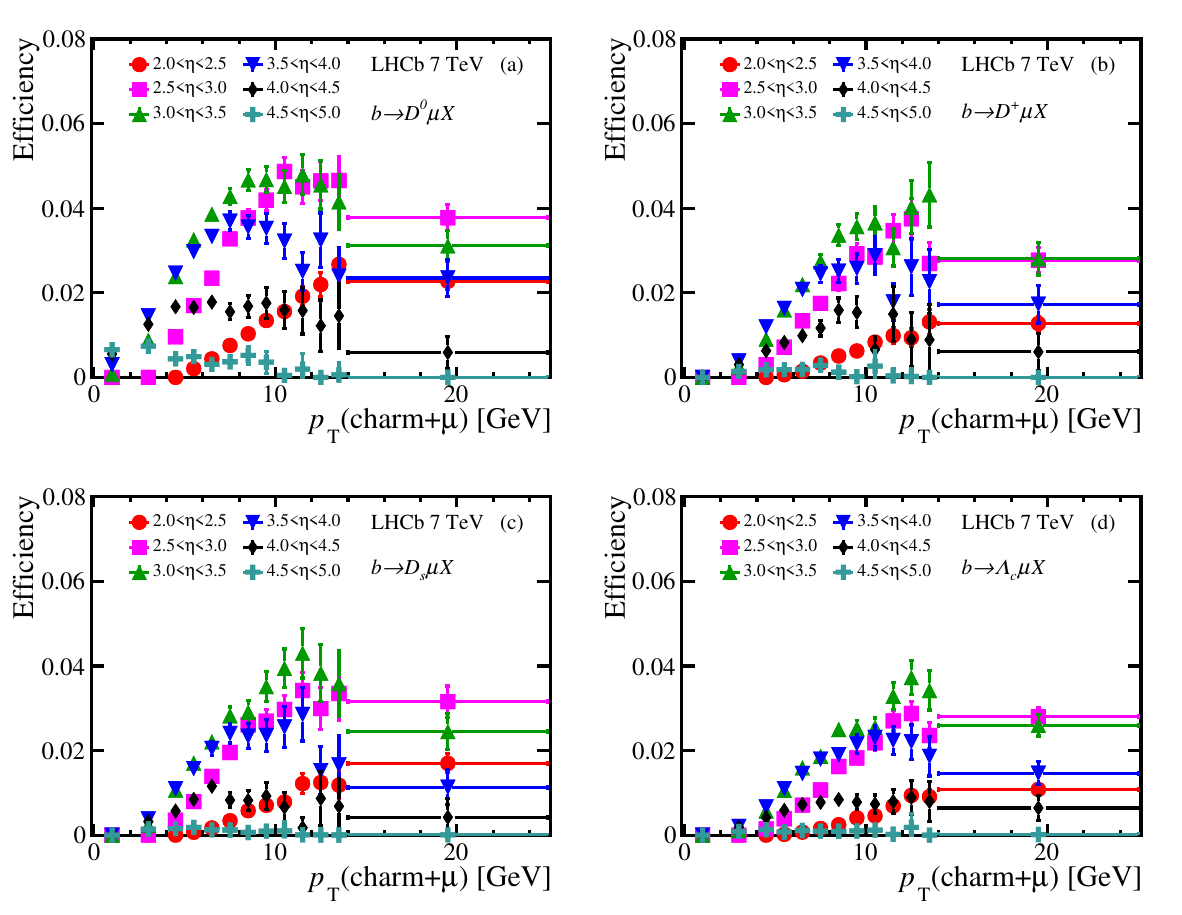}	
\caption{Overall detection efficiencies as a function of  \pt (charm+$\mu$) for the different $\eta$ intervals at 7~TeV. The uncertainties reflect both the statistical and systematic uncertainties added in quadrature. \label{pt_efficienciesFine7TeV}}
\end{figure}

We show here fits to the data for signal plus background integrated over $\eta$ for the various charm hadron plus muon combinations at 7~TeV in Fig.~\ref{fig:d0_example7} and 13~TeV in Fig.~\ref{fig:d0_example13}. We also show the overall detection efficiencies at 7~TeV in Fig.~\ref{pt_efficienciesFine7TeV}.

\clearpage
\newpage
\centerline{\large\bf LHCb collaboration}
\begin{flushleft}
\small
R.~Aaij$^{40}$,
B.~Adeva$^{39}$,
M.~Adinolfi$^{48}$,
Z.~Ajaltouni$^{5}$,
S.~Akar$^{6}$,
J.~Albrecht$^{10}$,
F.~Alessio$^{40}$,
M.~Alexander$^{53}$,
S.~Ali$^{43}$,
G.~Alkhazov$^{31}$
% ...

{\footnotesize \it
% ...
$ ^{5}$Clermont Universit{\'e}, Universit{\'e} Blaise Pascal, CNRS/IN2P3, LPC, Clermont-Ferrand, France\\
$ ^{6}$CPPM, Aix-Marseille Universit{\'e}, CNRS/IN2P3, Marseille, France\\
% ...
$ ^{10}$Fakult{\"a}t Physik, Technische Universit{\"a}t Dortmund, Dortmund, Germany\\
% ...
$ ^{31}$Petersburg Nuclear Physics Institute (PNPI), Gatchina, Russia\\
% ...
$ ^{39}$Universidad de Santiago de Compostela, Santiago de Compostela, Spain\\
$ ^{40}$European Organization for Nuclear Research (CERN), Geneva, Switzerland\\
% ...
$ ^{43}$Nikhef National Institute for Subatomic Physics, Amsterdam, The Netherlands\\
% ...
$ ^{48}$H.H. Wills Physics Laboratory, University of Bristol, Bristol, United Kingdom\\
% ...
$ ^{53}$School of Physics and Astronomy, University of Glasgow, Glasgow, United Kingdom\\
% ...
}
\end{flushleft}

\end{document}